\documentclass[twocolumn,amsmath,amssymb]{revtex4-1}
\usepackage{graphicx}
\usepackage{amssymb}
\usepackage{amsmath}
\usepackage{subfigure}
\usepackage{graphpap}
\usepackage{dcolumn} 
\usepackage{bm}
\usepackage{color,calc}
\usepackage{longtable}
\usepackage{color}
\usepackage{tikz}

\usepackage{multirow}
\usepackage{hyperref}
\renewcommand{\c}{{\rm c}}

\newcommand{\x}{\ensuremath{\text{x}}}
\renewcommand{\c}{\ensuremath{\text{c}}}

\newcommand{\bra}[1]{\ensuremath{\langle #1 \vert}}
\newcommand{\ket}[1]{\ensuremath{\vert #1  \rangle}}
\newcommand{\braket}[2]{\ensuremath{\langle  #1 \vert #2  \rangle}}

\renewcommand{\b}[1]{\ensuremath{\mathbf{#1}}}
\renewcommand{\l}{\ensuremath{\lambda}}
\newcommand{\s}{\ensuremath{\sigma}}

\renewcommand{\b}[1]{\ensuremath{\mathbf{#1}}}

\renewcommand{\s}{\text{s}}
\renewcommand{\d}{\text{d}}
\begin{document}

\preprint{}
\title{Self-consistent double-hybrid density-functional theory using the optimized-effective-potential method}
\author{Szymon \'Smiga$^{1,2,3}$}\email{szsmiga@fizyka.umk.pl}
\author{Odile Franck$^{4,5}$}\email{odile.franck@etu.upmc.fr}
\author{Bastien Mussard$^{4,5}$}\email{bastien.mussard@upmc.fr}
\author{Adam Buksztel$^1$}\email{abuk@fizyka.umk.pl}
\author{Ireneusz Grabowski$^1$}\email{ig@fizyka.umk.pl}
\author{Eleonora Luppi$^{4}$}\email{eleonora.luppi@upmc.fr}
\author{Julien Toulouse$^{4}$}\email{julien.toulouse@upmc.fr}
\affiliation{
$^1$Institute of Physics, Faculty of Physics, Astronomy and Informatics, Nicolaus Copernicus University, 87-100 Torun, Poland\\
$^2$Istituto Nanoscienze-CNR, Via per Arnesano 16, I-73100 Lecce, Italy\\
$^3$Center for Biomolecular Nanotechnologies @UNILE, Istituto Italiano di Tecnologia (IIT), Via Barsanti, 73010 Arnesano (LE), Italy\\
$^4$Laboratoire de Chimie Th\'eorique, Universit\'e Pierre et Marie Curie, CNRS, Sorbonne Universit\'es, F-75005 Paris, France\\
$^5$Institut des sciences du calcul et des donn\'ees, Universit\'e Pierre et Marie Curie, Sorbonne Universit\'es, F-75005, Paris, France
}

\date{September 20, 2016}
\begin{abstract}
We introduce an orbital-optimized double-hybrid (DH) scheme using the optimized-effective-potential (OEP) method. The orbitals are optimized using a local potential corresponding to the complete exchange-correlation energy expression including the second-order M{\o}ller-Plesset (MP2) correlation contribution. We have implemented a one-parameter version of this OEP-based self-consistent DH scheme using the BLYP density-functional approximation and compared it to the corresponding non-self-consistent DH scheme for calculations on a few closed-shell atoms and molecules. While the OEP-based self-consistency does not provide any improvement for the calculations of ground-state total energies and ionization potentials, it does improve the accuracy of electron affinities and restores the meaning of the LUMO orbital energy as being connected to a neutral excitation energy. Moreover, the OEP-based self-consistent DH scheme provides reasonably accurate exchange-correlation potentials and correlated densities. 
\end{abstract}

\maketitle

\section{Introduction}

Density-functional theory (DFT)~\cite{HohKoh-PR-64,KohSha-PR-65} is a powerful approach for electronic-structure calculations of atoms, molecules, and solids. In the Kohn-Sham (KS) formulation, series of approximations for the exchange-correlation energy have been developed for an ever-increasing accuracy: local-density approximation (LDA), semilocal approximations (generalized-gradient approximations (GGA) and meta-GGA), hybrid approximations introducing a fraction of Hartree-Fock (HF) exchange, and nonlocal correlation approximations using virtual KS orbitals (see, e.g., Ref.~\onlinecite{Bec-JCP-14} for a recent review).

In the latter family of approximations, the double-hybrid (DH) approximations are becoming increasingly popular. Introduced in their current form by Grimme~\cite{Gri-JCP-06}, they consist in combining a semilocal exchange density functional with HF exchange and a semilocal correlation density functional with second-order M{\o}ller-Plesset (MP2) perturbative correlation. Numerous such DH approximations have been developed in the last decade (see Ref.~\onlinecite{GoeGri-WIRE-14} for a review). In general, DH approximations give thermochemistry properties with near-chemical accuracy for molecular systems without important static correlation effects.
In virtually all applications of DH approximations, the orbitals are calculated within the generalized KS (GKS) framework~\cite{SeiGorVogMajLev-PRB-96} (i.e., with a nonlocal HF exchange potential) and without the presence of the MP2 correlation term. The MP2 contribution is then evaluated using the previously self-consistently calculated orbitals and added a posteriori to the total energy.
Recently, Peverati and Head-Gordon~\cite{PevHea-JCP-13} proposed an orbital-optimized DH scheme where the orbitals are self-consistently optimized in the presence of the MP2 correlation term. This is a direct extension of orbital-optimized MP2 schemes in which the MP2 total energy is minimized with respect to occupied-virtual orbital rotation parameters~\cite{LocHea-JCP-07,NeeSchKosSchGri-JCTC-09}. Like for regular orbital-optimized MP2 method, also here the optimization of orbitals leads to substantial improvements in spin-unrestricted calculations for symmetry breaking and open-shell situations. Very recently, an approximate orbital-optimized DH scheme was also proposed which confirmed the utility of optimizing the orbitals in complicated electronic-structure problems~\cite{SanPerSavBreAda-JPCA-16}.

In this work, we propose an alternative orbital-optimized DH scheme using the optimized-effective-potential (OEP) method~\cite{ShaHor-PR-53,TalSha-PRA-76} (see Refs.~\onlinecite{GraKreKurGro-INC-00,Eng-INC-03,KumKro-RMP-08} for reviews on OEP). The idea is to optimize the orbitals in the DH total energy expression by using a fully local potential corresponding to the complete exchange-correlation energy expression including the MP2 contribution. This can be considered as an extension of OEP schemes using a second-order correlation energy expression~\cite{GorLev-PRA-94,GorLev-IJQC-95,GraHirIvaBar-JCP-02,BarGraHirIva-JCP-05,MorWuYan-JCP-05}.

In comparison to the previously-mentioned orbital-optimized DH schemes, we expect that the proposed OEP self-consistent DH scheme to provide additional advantages. First, there is the appeal of staying within the philosophy of the KS scheme with a local potential. Second, having a local potential and the associated orbital energies can be useful for interpretative purposes. Third, the OEP approach can be advantageous in calculations of excitation energies and response properties, similarly to the advantages of using OEP exact exchange (EXX) versus regular HF. Indeed, contrary to the HF case, with EXX the unoccupied orbitals feel a local potential asymptotically decaying as $-1/r$, allowing it to support many unoccupied bound states, which are good starting points for calculating high-lying/Rydberg excitation energies and response properties (see, e.g., Refs.~\onlinecite{GraKreKurGro-INC-00,HirIvaGraBar-JCP-02,KumKro-RMP-08}).

The paper is organized as follows. In Section~\ref{sec:theory}, we review the theory of the standard DH approximations and formulate the proposed self-consistent OEP DH approach. We also explain how the ionization potential and the electron affinity are obtained in both methods. After providing computational details in Section~\ref{sec:comput}, we discuss our results in Section~\ref{sec:results} on total energies, ionization potentials, electron affinities, exchange-correlation and correlation potentials, and correlated densities obtained for a set of atoms (He, Be, Ne, Ar) and molecules (CO and H$_2$O). Finally, Section~\ref{sec:conclusion} contains our conclusions.

Throughout the paper, we use the convention that $i$ and $j$ indices label occupied spin orbitals, $a$ and $b$ label virtual ones, and $p$ and $q$ are used for both occupied and virtual spin orbitals. In all equations Hartree atomic units are assumed.

\section{Theory}
\label{sec:theory}

\subsection{Standard double-hybrid approximations}

For simplicity, in this work, we consider the one-parameter double-hybrid (1DH) approximation of Ref.~\onlinecite{ShaTouSav-JCP-11} in which the density scaling in the correlation functional is neglected.
The extension to more general density-scaled or two-parameter double-hybrid approximations~\cite{Gri-JCP-06} is straightforward. The expression of the total energy is thus written as
\begin{eqnarray}
E &=& \sum_i
\int \varphi_i^*(\b{x}) \left( -\frac{1}{2} \bm{\nabla}^2 +v_{\text{ne}}(\b{r}) \right) \varphi_i(\b{x}) \; \d \b{x}
\nonumber\\
&&+ E_\text{H} + E_\text{xc}^{\text{1DH}},
\label{E1DH}
\end{eqnarray}
 where $\varphi_i(\b{x})$  are the occupied spin orbitals with $\b{x}=(\b{r},\sigma)$ indicating
 space-spin coordinates. In Eq.~(\ref{E1DH}), $v_{\text{ne}}(\b{r})$ is the nuclei-electron potential, $E_\text{H} = (1/2) \iint n(\b{x}_1) n(\b{x}_2)/|\b{r}_2-\b{r}_1| \d \b{x}_1 \d \b{x}_2$ is the Hartree energy written 
with the spin densities $n(\b{x}) = \sum_i
|\varphi_i(\b{x})|^2$, and $E_\text{xc}^{\text{1DH}}$ is the exchange-correlation energy taken as
\begin{equation}
E_\text{xc}^{\text{1DH}} = E_\text{xc}^{\text{1H}} + \lambda^2 \;E_\text{c}^\text{MP2}.
\label{Exc1DH}
\end{equation}
In this expression, $E_\text{xc}^{\text{1H}}$ is the one-parameter hybrid (1H) part of the exchange-correlation energy
\begin{equation}
E_\text{xc}^{\text{1H}} = \lambda E_\text{x}^\text{HF} + (1-\lambda) E_\text{x}^\text{DFA} + (1-\lambda^2) E_\text{c}^\text{DFA},
\label{Exc1H}
\end{equation}
and $\lambda$ ($0 \leqslant \lambda \leqslant 1$) is an empirical scaling parameter. The expression of the HF (or exact) exchange energy is
\begin{equation}
E_\text{x}^{\text{HF}} = - \frac{1}{2} \sum_{i,j}
\braket{ij}{ji},
\end{equation}
where $\braket{pq}{rs}=\iint \d \b{x}_1 \d \b{x}_2 \varphi_p^*(\b{x}_1) \varphi_q^*(\b{x}_2) \varphi_r(\b{x}_1) \varphi_s(\b{x}_2)/|\b{r}_2-\b{r}_1|$ are the two-electron integrals. The expression for the MP2 correlation energy is
\begin{equation}
E_\text{c}^{\text{MP2}} = - \frac{1}{4} \sum_{i,j}
\sum_{a,b}
 \frac{|\bra{ij}\ket{ab}|^2}{\varepsilon_a + \varepsilon_b - \varepsilon_i - \varepsilon_j},
\label{EcMP2}
\end{equation}
where $\bra{ij}\ket{ab}=\braket{ij}{ab} - \braket{ij}{ba}$ are the antisymmetrized two-electron integrals, and $\varepsilon_p$ is the energy of the spin orbital $p$. Finally, $E_\text{x}^\text{DFA}$ and $E_\text{c}^\text{DFA}$ are the 
semilocal density-functional approximations (DFA) evaluated at the spin densities $n(\b{x})$. For example, choosing the Becke 88 (B) exchange functional~\cite{Bec-PRA-88} and the Lee-Yang-Parr (LYP) correlation functional~\cite{LeeYanPar-PRB-88} leads to the 1DH-BLYP double-hybrid approximation~\cite{ShaTouSav-JCP-11} which is a one-parameter version of the B2-PLYP approximation~\cite{Gri-JCP-06}.

We must stress here that the expression of the correlation energy in Eq.~(\ref{EcMP2}) has a standard MP2 form. However, except for $\lambda = 1$, the orbitals are not the HF ones, so the value of the correlation energy in Eq.~(\ref{EcMP2}) calculated with these orbitals does not correspond to the standard MP2 correlation energy. In the DFT context, the most usual second-order correlation energy expression is given by second-order G\"orling-Levy (GL2) perturbation theory~\cite{GorLev-PRA-94,GorLev-IJQC-95}, which in addition to the MP2-like double-excitation term also includes a single-excitation term. In this work, following the standard practice for the double-hybrid approximations, we do not include the single-excitation term, which is usually two orders of magnitude smaller than the double-excitation term~\cite{grabowski:2007:ccpt2,grabowski:2008:ijqc,GraFabTeaSmiBukDel-JCP-14}.

In the standard DH approximations, the spin orbitals are calculated by disregarding the MP2 term in Eq.~(\ref{Exc1DH}) and considering the HF exchange energy as a functional of the one-particle density matrix
$n_1(\b{x}',\b{x})=\sum_i
\varphi_i^*(\b{x}) \varphi_i(\b{x}')$, leading to a
GKS equation
\begin{eqnarray}
\left( -\frac{1}{2} \bm{\nabla}^2 +v_{\text{ne}}(\b{r}) + v_\text{H}(\b{r}) \right) \varphi_p(\b{x}) 
\nonumber\\
+ \int v_{\text{xc}}^{\text{1H}}(\b{x},\b{x}') \varphi_p(\b{x}') \d \b{x}' = \varepsilon_p^\text{1H} \varphi_p(\b{x}),
\label{hnlpsi}
\end{eqnarray}
where $v_\text{H}(\b{r}) = \int n(\b{x}')/|\b{r}'-\b{r}| \d \b{x}'$ is the Hartree potential and $v_{\text{xc}}^{\text{1H}}(\b{x},\b{x}')$ is the functional derivative of the three terms in Eq.~(\ref{Exc1H}) with respect to $n_1(\b{x}',\b{x})$
\begin{eqnarray}
v_{\text{xc}}^{\text{1H}}(\b{x},\b{x}') &=& \frac{\delta E_\text{xc}^{\text{1H}}}{\delta n_1(\b{x}',\b{x})}
\nonumber\\
&=&\l v_{\text{x}}^{\text{HF}}(\b{x},\b{x}') + (1-\l) v_{\text{x}}^{\text{DFA}}(\b{x}) \delta(\b{x}-\b{x}')
\nonumber\\
&& + (1-\l^2) v_{\text{c}}^{\text{DFA}}(\b{x}) \delta(\b{x}-\b{x}'). \;
\label{vxc1H}
\end{eqnarray}
In this expression, $v_{\text{x}}^{\text{HF}}(\b{x},\b{x}') = - n_1(\b{x},\b{x}')/|\b{r}-\b{r}'|$ is the nonlocal HF potential, while $v_{\text{x}}^{\text{DFA}}(\b{x}) = \delta E_\text{x}^\text{DFA} /\delta n(\b{x})$ and $v_{\text{c}}^{\text{DFA}}(\b{x}) = \delta E_\text{c}^\text{DFA} /\delta n(\b{x})$ are the local exchange and correlation DFA potentials, respectively. These 1H orbitals and corresponding orbital energies $\varepsilon_p^\text{1H}$ are thus used in the MP2 correlation expression of Eq.~(\ref{EcMP2}). We recall that for $\lambda=0$ the 1DH method reduces to the standard KS scheme, while for $\lambda=1$ it recovers the standard MP2 method with HF orbitals. In practice, optimal values of $\lambda$ are around 0.6-0.8, depending on the density-functional approximations used~\cite{ShaTouSav-JCP-11,SouShaTou-JCP-14,ShaTouMasCiv-JCP-14}.

\subsection{Self-consistent OEP double-hybrid approximations}

Here, we propose to fully self-consistently calculate the spin orbitals in the DH approximations by taking into account the MP2 term, and considering the HF exchange energy and MP2 correlation energy as implicit functionals of the density. Thus, Eq.~(\ref{hnlpsi}) is replaced by a KS equation
\begin{eqnarray}
\left( -\frac{1}{2} \bm{\nabla}^2 +v_{\text{ne}}(\b{r}) + v_\text{H}(\b{r}) +  v_{\text{xc}}^{\text{OEP-1DH}}(\b{x})\right) \varphi_p(\b{x}) 
\nonumber\\
= \varepsilon_p \varphi_p(\b{x}),
\label{hOEPpsi}
\end{eqnarray}
where $v_{\text{xc}}^{\text{OEP-1DH}}$ is a fully local potential obtained by taking the functional derivative with respect to the density of all terms in Eq.~(\ref{Exc1DH})
\begin{eqnarray}
v_{\text{xc}}^{\text{OEP-1DH}}(\b{x}) &=& \frac{\delta E_\text{xc}^{\text{1DH}}}{\delta n(\b{x})}
\nonumber\\
&=&\l v_{\text{x}}^{\text{EXX}}(\b{x}) + (1-\l) v_{\text{x}}^{\text{DFA}}(\b{x})
\nonumber\\
&&+ (1-\l^2) v_{\text{c}}^{\text{DFA}}(\b{x}) + \l^2 v_{\text{c}}^{\text{GL2}}(\b{x}), \;
\label{vxcOEP1DH}
\end{eqnarray}
where $v_{\text{x}}^{\text{EXX}}(\b{x}) = \delta E_\text{x}^\text{HF}/\delta n(\b{x})$ is the EXX potential and $v_{\text{c}}^{\text{GL2}}(\b{x}) = \delta E_\text{c}^\text{MP2}/\delta n(\b{x})$ is here referred to as the GL2 correlation potential (even though it does not contain the single-excitation term). Since $E_\text{x}^\text{HF}$ and $E_\text{c}^\text{MP2}$ are only implicit functionals of the density through the orbitals and orbital energies, the calculation of $v_{\text{x}}^{\text{EXX}}(\b{x})$ and $v_{\text{c}}^{\text{GL2}}(\b{x})$ must be done with the OEP method, as done in Refs.~\onlinecite{GraHirIvaBar-JCP-02,MorWuYan-JCP-05,BarGraHirIva-JCP-05}. We note that several alternative methods to OEP have been proposed~\cite{KriLiIaf-PRA-92,Cas-PRA-95,DelGor-JCP-01,GruGriBae-JCP-02,Hes-PCCP-06,FabDel-JCP-07,IzmStaScuDav-JCP-07}, but we do not consider these alternative methods in this work. We will refer to the present approach as the OEP-1DH method. As in the case of the standard DH approach, for $\lambda=0$ the OEP-1DH method reduces to the standard KS scheme. For $\lambda=1$ it reduces to a correlated OEP scheme with the full MP2-like correlation energy expression (but without the single-excitation term), here referred to as the OEP-GL2 scheme.

The OEP equations for the EXX exchange and GL2 correlation potentials 
\begin{equation}
\int v_{\x}^{\text{EXX}}(\b{x}') \; \chi_\s (\b{x}',\b{x}) \; \d \b{x}' = \Lambda_\x (\b{x}),
\label{vxEXX}
\end{equation}
and
\begin{equation}
\int v_{\c}^{\text{GL2}}(\b{x}') \; \chi_\s (\b{x}',\b{x}) \; \d \b{x}' = \Lambda_\c^{\text{MP2}}(\b{x}),
\label{vcMP2}
\end{equation}
can be obtained after applying a functional-derivative chain rule (see, e.g., Refs.~\onlinecite{EngDre-JCC-99,GraHirIvaBar-JCP-02,Eng-INC-03,MorWuYan-JCP-05,GraFabTeaSmiBukDel-JCP-14}). In these expressions, $\chi_\s (\b{x}',\b{x})=\delta n(\b{x}')/\delta v_\s(\b{x})$  is the  KS static linear-response function which can be expressed in terms of spin orbitals and spin orbital energies,
\begin{equation}
\chi_\s(\b{x}',\b{x}) = - \sum_i
\sum_a
\frac{\varphi_i^*(\b{x}')\varphi_a(\b{x}') \varphi_a^*(\b{x})\varphi_i(\b{x}) }{\varepsilon_a - \varepsilon_i} + \text{c.c.},
\end{equation}
where c.c. stands for the complex conjugate, and $v_\s(\b{x}) = v_{\text{ne}}(\b{r}) + v_\text{H}(\b{r}) +  v_{\text{xc}}^{\text{OEP-1DH}}(\b{x})$ is the total KS potential. The expressions for $\Lambda_\x (\b{x})$ and  $\Lambda_\c^{\text{MP2}} (\b{x})$ are, respectively,
\begin{eqnarray}
\Lambda_\x (\b{x}) &=& \frac{\delta E_\x^\text{HF}}{\delta v_\s(\b{x})}
= \sum_i
\int \d \b{x}' \left( \frac{\delta E_\x^\text{HF}}{\delta \varphi_i(\b{x}')} \frac{\delta \varphi_i(\b{x}')}{\delta v_\s(\b{x})} + \text{c.c.} \right)
\nonumber\\
&=& \sum_{i,j}
 \sum_a
\left( \braket{ij}{ja} \frac{\varphi_a^*(\b{x}) \varphi_i(\b{x})}{\varepsilon_a -\varepsilon_i} + \text{c.c.}\right),
\end{eqnarray}
and
\begin{widetext}
\begin{eqnarray}
\Lambda_\c^{\text{MP2}} (\b{x}) &=& \frac{\delta E_\c^\text{MP2}}{\delta v_\s(\b{x})}
= \sum_p
 \Biggl[ \int \d \b{x}' \left( \frac{\delta E_\c^\text{MP2}}{\delta \varphi_p(\b{x}')} \frac{\delta \varphi_p(\b{x}')}{\delta v_\s(\b{x})} + \text{c.c.} \right)
+ \frac{\partial E_\c^\text{MP2}}{\partial \varepsilon_p} \frac{\delta \varepsilon_p}{\delta v_\s(\b{x})} \Biggl]
\nonumber\\
&=& \frac{1}{2} \sum_{i,j}
\sum_{a,b}
 \sum_{q\not=i}
\left( \frac{\bra{ij}\ket{ab}\bra{ab}\ket{qj}}{\varepsilon_a+\varepsilon_b-\varepsilon_i-\varepsilon_j} \frac{\varphi_q^*(\b{x})\varphi_i(\b{x})}{\varepsilon_q -\varepsilon_i} + \text{c.c.} \right)
+ \frac{1}{2} \sum_{i,j}
 \sum_{a,b}
 \sum_{q\not=a}
 \left( \frac{\bra{ij}\ket{qb}\bra{ab}\ket{ij}}{\varepsilon_a+\varepsilon_b-\varepsilon_i-\varepsilon_j} \frac{\varphi_q^*(\b{x})\varphi_a(\b{x})}{\varepsilon_q -\varepsilon_a} + \text{c.c.} \right)
\nonumber\\
&&+ \frac{1}{2} \sum_{i,j}
 \sum_{a,b}
 \frac{|\bra{ij}\ket{ab}|^2}{(\varepsilon_a + \varepsilon_b - \varepsilon_i - \varepsilon_j)^2} \left( |\varphi_a(\b{x})|^2 - |\varphi_i(\b{x})|^2 \right).
\end{eqnarray}
\end{widetext}

In practice, in order to solve the OEP equations [Eqs.~(\ref{vxEXX}) and~(\ref{vcMP2})], the EXX and GL2 potentials are calculated using expansions in a finite Gaussian basis set~\cite{IvaHirBar-PRL-99,Gor-PRL-99,HirIvaGraBarBurTal-JCP-01,IvaHirBar-JCP-02,YanWu-PRL-02}.
 The EXX potential is thus expanded over orthonormalized auxiliary Gaussian basis functions $\{g_n(\b{r})\}$ as
\begin{equation}
v_{\x}^{\text{EXX}}(\b{r}\sigma) = v_\text{Slater}(\b{r}\sigma) + \sum_{n} c_{\x,n}^\sigma \; g_n(\b{r}),
\label{vxEXXbasis}
\end{equation}
where the Slater potential, $v_\text{Slater}(\b{r}\sigma) = -(1/n(\b{x}))\int \d \b{x}' |n_1(\b{x},\b{x}')|^2/|\b{r}-\b{r}'|$, is added to ensure the correct $-1/|\b{r}|$ asymptotic behavior of the potential. Similarly, the GL2 potential is expanded as
\begin{equation}
v_{\c}^{\text{GL2}}(\b{r}\sigma) = \sum_{n} c_{\c,n}^\sigma \; g_n(\b{r}).
\label{vcMP2basis}
\end{equation}
Expanding as well the linear-response function in the same basis
\begin{equation}
\chi_\s (\b{r}\sigma,\b{r}'\sigma) = \sum_{n,m} (\b{X}_\sigma)_{nm} g_n(\b{r}) g_m(\b{r}'),
\end{equation}
and after using Eq.~(\ref{vxEXX}), the coefficients in Eq.~(\ref{vxEXXbasis}) are found as
\begin{equation} \label{cx}
c_{\x,n}^\sigma = \sum_{m} (\bm{\Lambda}_{\x,\sigma})_{m}  (\b{X}_\sigma^{-1})_{mn} - v_{\text{Slater},n}^\sigma,
\end{equation}
where $(\bm{\Lambda}_{\x,\sigma})_{m} = \int \d\b{r} \; g_m(\b{r}) \Lambda_\x (\b{r}\sigma)$, $v_{\text{Slater},n}^\sigma = \int \d \b{r} \; g_n(\b{r}) v_\text{Slater}(\b{r}\sigma)$, and $(\b{X}_\sigma^{-1})_{mn}$ are the elements of the (pseudo-)inverse of the matrix $\b{X}_\sigma$. Similarly, after using Eq.~(\ref{vcMP2}), the coefficients in Eq.~(\ref{vcMP2basis}) are found as
\begin{equation} \label{cc}
c_{\c,n}^\sigma = \sum_{m} (\bm{\Lambda}^\text{MP2}_{\c,\sigma})_{m}  (\b{X}_\sigma^{-1})_{mn},
\end{equation}
where $(\bm{\Lambda}^\text{MP2}_{\c,\sigma})_{m} = \int \d\b{r} \; g_m(\b{r}) \Lambda_\c^{\text{MP2}} (\b{r}\sigma)$. In this work, the same basis set is used for expanding the orbitals and the potentials. 
In practice, our OEP-1DH calculations employ a truncated singular-value decomposition (TSVD) method for the construction of the pseudo-inverse of the linear-response function [used in Eqs.~(\ref{cx}) and~(\ref{cc})] to ensure that stable and physically sound solutions are obtained in the OEP equations [Eqs.~(\ref{vxEXX}) and~(\ref{vcMP2})].

In principle, this procedure selects the EXX and GL2 potentials which vanish at $|\b{r}|\to\infty$. We note that, when continuum states are included, the GL2 potential actually diverges at infinity for finite systems~\cite{FacEngSchDre-PRL-01,NiqFucGon-PRL-03,FacEngSchDre-PRL-03,IvaLev-JCP-02,EngJiaFac-PRA-05}. Nevertheless, this problem is avoided when using a discrete basis set with functions vanishing at infinity (such as the basis set used in this work)~\cite{IvaLev-JCP-02,NiqFucGon-JCP-03,EngJiaFac-PRA-05}. In practice, the calculated potentials can still be shifted by a function which vanishes at infinity but which is an arbitrary constant in the physically relevant region of space. To remove this arbitrary constant, as in Ref.~\onlinecite{GraHirIvaBar-JCP-02}, we impose the HOMO condition on the EXX potential
\begin{eqnarray}
v_{\x,HH}^\text{EXX} = v_{\x,HH}^\text{HF},
\label{HOMOcondEXX}
\end{eqnarray}
where $v_{\x,HH}^\text{EXX}=\int \varphi_H^*(\b{x}) v_\x^\text{EXX}(\b{x}) \varphi_H(\b{x}) \d\b{x}$ and $v_{\x,HH}^\text{HF}=\iint \varphi_H^*(\b{x}) v_\x^\text{HF}(\b{x},\b{x}') \varphi_H(\b{x}') \d\b{x}\d\b{x}'=-
\sum_j
 \braket{Hj}{jH}$ are the expectation values of the EXX and HF exchange potentials over the HOMO spin orbital referred to as $H$. Similarly, we impose the HOMO condition on the GL2 potential
\begin{eqnarray}
v_{\c,HH}^\text{GL2}  = \Sigma^\text{MP2}_{\c,HH}(\varepsilon_H),
\label{HOMOcondMP2}
\end{eqnarray}
where $v_{\c,HH}^\text{GL2}=\int \varphi_H^*(\b{x}) v_\c^\text{GL2}(\b{x}) \varphi_H(\b{x}) \d\b{x}$ and $\Sigma^\text{MP2}_{\c,HH}(\varepsilon_H)=\iint \varphi_H^*(\b{x}) \Sigma_\text{c}^\text{MP2}(\b{x},\b{x}';\varepsilon_H) \varphi_H(\b{x}') \d\b{x}\d\b{x}'$ are the expectation values of the GL2 local potential and of the MP2 self-energy over the HOMO spin orbital. The MP2 self-energy is defined as the functional derivative of $E_\text{c}^\text{MP2}$ with respect to the one-particle Green function $G(\b{x}',\b{x};\omega)$, i.e. $\Sigma_\text{c}^\text{MP2}(\b{x},\b{x}';\omega) = 2\pi i \; \delta E_\text{c}^\text{MP2}/\delta G(\b{x}',\b{x};\omega)$, and its diagonal matrix elements $\Sigma^\text{MP2}_{\c,pp}(\omega)$ are~\cite{SzaOst-BOOK-96}
\begin{eqnarray}
\Sigma^\text{MP2}_{\c,pp}(\omega) &=& 
-\frac{1}{2} \sum_{j}
 \sum_{a,b}
  \frac{|\bra{pj}\ket{ab}|^2}{\varepsilon_a+\varepsilon_b-\omega-\varepsilon_j}
\nonumber\\
&& + \frac{1}{2} \sum_{i,j}
 \sum_{b}
 \frac{|\bra{ij}\ket{pb}|^2}{\omega+\varepsilon_b-\varepsilon_i-\varepsilon_j}.
\label{SigmaMP2}
\end{eqnarray}
Eq.~(\ref{HOMOcondEXX}) can be obtained either by considering the asymptotic limit of Eq.~(\ref{vxEXX}) and using the fact that the HOMO spin orbital dominates in this limit over all occupied spin orbitals~\cite{KriLiIaf-PRA-92a}, or by considering the derivative of the HF exchange energy with respect to the electron number (at fixed potential, i.e. at fixed orbitals) and using the chain rule with either the one-particle density or the one-particle density matrix~\cite{KriLiIaf-PRA-92a,YanCohMor-JCP-12}. Similarly, Eq.~(\ref{HOMOcondMP2}) can be obtained by considering the derivative of the MP2 correlation energy with respect to the electron number (at fixed potential) and using the chain rule with either the one-particle density or the one-particle Green function~\cite{Cas-PRB-99}. For systems with degenerate HOMO orbitals, we introduce in Eqs.~(\ref{HOMOcondEXX}) and~(\ref{HOMOcondMP2}) sums over the degenerate HOMOs divided by the number of such HOMOs, $(1/n_H) \sum_H$, as done in Ref.~\onlinecite{HesGotDelGor-JCP-07}.

\subsection{Ionization potential and electronic affinity}

The ionization potential (IP) and the electronic affinity (EA) can be defined as derivatives of the total energy with respect to the electron number $N$. For the self-consistent OEP DH approximations, these derivatives can be expressed in terms of frontier spin orbital energies, like in exact KS DFT,~\cite{PerParLevBal-PRL-82,Cas-PRB-99,CohMorYan-PRB-08,YanCohMor-JCP-12}
\begin{equation}
-\text{IP}^{\text{OEP-1DH}} = \left(\frac{\partial E^{\text{OEP-1DH}}}{\partial N} \right)_{N-\delta}  = \varepsilon_H,
\label{IPOEP1DH}
\end{equation}
and
\begin{equation}
-\text{EA}^{\text{OEP-1DH}} = \left(\frac{\partial E^{\text{OEP-1DH}}}{\partial N} \right)_{N+\delta}  = \varepsilon_L + \Delta_\text{xc},
\label{EAOEP1DH}
\end{equation}
where $\delta \to 0^+$, $L$ refers to the LUMO spin orbital, and $\Delta_\text{xc}$ is the derivative discontinuity of the exchange-correlation energy. For the OEP-1DH approximation, the derivative discontinuity comes from the $\lambda$-scaled EXX and GL2 contributions
\begin{equation}
\Delta_\text{xc} = \l \; \Delta_\text{x}^{\text{EXX}} + \l^2 \;
\Delta_\text{c}^{\text{GL2}}.
\end{equation}
The terms $\Delta_\text{x}^{\text{EXX}}$ and $\Delta_\text{c}^{\text{GL2}}$ are given by~\cite{GorLev-PRA-95}
\begin{eqnarray}
\Delta_\text{x}^{\text{EXX}} =  \left( v_{\x,LL}^\text{HF} - v_{\x,LL}^\text{EXX} \right) - \left( v_{\x,HH}^\text{HF} - v_{\x,HH}^\text{EXX} \right),
\label{Deltax}
\end{eqnarray}
where $v_{\x,LL}^\text{HF}=\iint \varphi_L^*(\b{x}) v_\x^\text{HF}(\b{x},\b{x}') \varphi_L(\b{x}') \d\b{x}\d\b{x}'=-
\sum_j
\braket{Lj}{jL}$ and $v_{\x,LL}^\text{EXX}=\int \varphi_L^*(\b{x}) v_\x^\text{EXX}(\b{x}) \varphi_L(\b{x}) \d\b{x}$ are the expectation values of the HF and EXX exchange potentials over the LUMO spin orbital, and similarly~\cite{Cas-PRA-95,Cas-PRB-99}
\begin{eqnarray}
\Delta_\text{c}^{\text{GL2}} = \left( \Sigma_{\text{c},LL}^\text{MP2}(\varepsilon_L) - v_{\c,LL}^\text{GL2} \right) -  \left( \Sigma_{\text{c},HH}^\text{MP2}(\varepsilon_H) - v_{\c,HH}^\text{GL2} \right),
\nonumber\\
\label{Deltac}
\end{eqnarray}
where $\Sigma_{\text{c},LL}^\text{MP2}(\varepsilon_L)=\iint \varphi_L^*(\b{x}) \Sigma_\text{c}^\text{MP2}(\b{x},\b{x}';\varepsilon_L) \varphi_L(\b{x}') \d\b{x}\d\b{x}'$ and $v_{\c,LL}^\text{GL2}=\int \varphi_L^*(\b{x}) v_\c^\text{GL2}(\b{x}) \varphi_L(\b{x}) \d\b{x}$ are the expectation values of the MP2 self-energy and of the GL2 local potential over the LUMO spin orbital. Clearly, if the HOMO condition of Eqs.~(\ref{HOMOcondEXX}) and (\ref{HOMOcondMP2}) is imposed, then the differences of terms in the second parenthesis in Eqs.~(\ref{Deltax}) and~(\ref{Deltac}) are in fact zero. Note that Eq.~(\ref{Deltac}) can be found from the linearized version of the Sham-Schl\"uter equation~\cite{ShaSch-PRL-83}. Again, for degenerate HOMOs and/or LUMOs, we introduce in Eqs.~(\ref{Deltax}) and~(\ref{Deltac}), sums over the degenerate HOMOs/LUMOs divided by the number of such degenerate HOMOs/LUMOs, i.e. $(1/n_H) \sum_H$ and $(1/n_L) \sum_L$.

For standard DH approximations, following Refs.~\onlinecite{CohMorYan-JCTC-09,SuYanMorXu-JPCA-14}, we obtain IPs and EAs by calculating derivatives of the total energy by finite differences
\begin{equation}
-\text{IP}^{\text{1DH}} = \left(\frac{\partial E^{\text{1DH}}}{\partial N} \right)_{N-\delta}  \approx \frac{E^{\text{1DH}}(N) - E^{\text{1DH}}(N-\Delta)}{\Delta},
\label{IP1DH}
\end{equation}
and
\begin{equation}
-\text{EA}^\text{1DH} = \left(\frac{\partial E^{\text{1DH}}}{\partial N} \right)_{N+\delta}  \approx \frac{E^{\text{1DH}}(N+\Delta) - E^{\text{1DH}}(N)}{\Delta},
\label{EA1DH}
\end{equation}
with $\Delta = 0.001$. To calculate the energies for fractional electron numbers, $E^{\text{1DH}}(N-\Delta)$ and $E^{\text{1DH}}(N+\Delta)$, we use the extension of the DH total energy expression, including the MP2 correlation term, to fractional orbital occupation numbers, as given in Refs.~\onlinecite{CohMorYan-JCTC-09,SuYanMorXu-JPCA-14} (for the details of our implementation, see Ref.~\onlinecite{MusTou-MP-16}). As pointed out in Ref.~\onlinecite{CohMorYan-JCTC-09}, if the variation of the orbitals and orbital energies in the MP2 correlation energy is neglected when taking the derivative of the 1DH total energy with the respect to $N$, then Eqs.~(\ref{IP1DH}) and~(\ref{EA1DH}) simplify to
\begin{equation}
-\text{IP}^\text{1DH} = \left(\frac{\partial E^{\text{1DH}}}{\partial N} \right)_{N-\delta}  \approx \varepsilon_H^\text{1H} +   \Sigma_{\text{c},HH}^\text{MP2}(\varepsilon_H^\text{1H}),
\label{IP1DHapprox}
\end{equation}
and
\begin{equation}
-\text{EA}^\text{1DH} = \left(\frac{\partial E^{\text{1DH}}}{\partial N} \right)_{N+\delta}  \approx \varepsilon_L^\text{1H} + \Sigma_{\text{c},LL}^\text{MP2}(\varepsilon_L^\text{1H}),
\label{EA1DHapprox}
\end{equation}
which corresponds to standard second-order perturbative propagator theory (see, e.g., Ref.~\onlinecite{SzaOst-BOOK-96}). Even though in practice we calculate $\text{IP}^{\text{1DH}}$ and $\text{EA}^\text{1DH}$ using Eqs.~(\ref{IP1DH}) and~(\ref{EA1DH}), the approximate connection with the self-energy in Eqs.~(\ref{IP1DHapprox}) and~(\ref{EA1DHapprox}) is useful for comparison and interpretative purposes. For example, it can be shown that $\Sigma_{\text{c},HH}^\text{MP2}(\varepsilon_H^\text{1H})$ contains a term corresponding to orbital relaxation in the $(N-1)$-electron system, and pair-correlation terms for the $N$- and $(N-1)$-electron systems~\cite{PicGos-MP-73,SzaOst-BOOK-96}.

\section{Computational details}
\label{sec:comput}

\begin{figure*}
\includegraphics[scale=0.45]{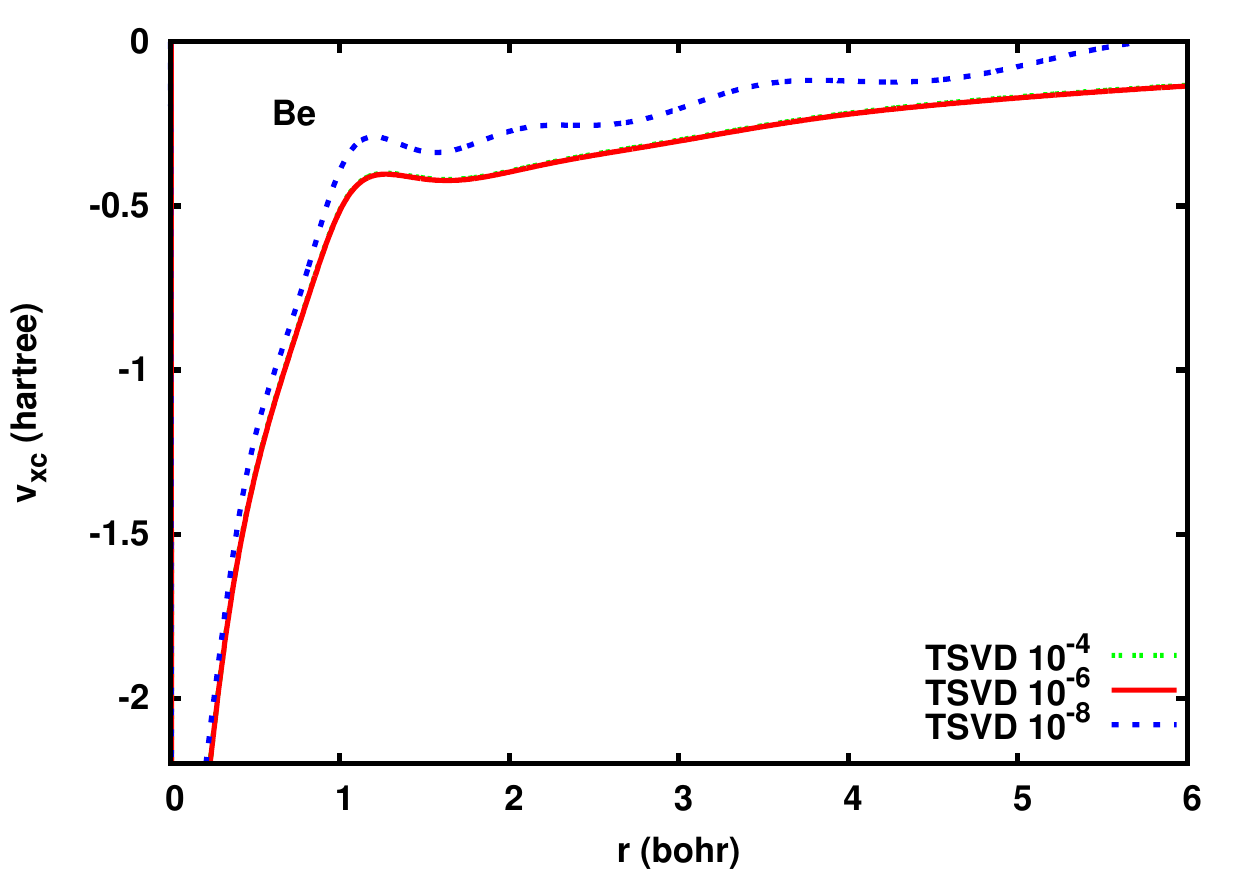}
\includegraphics[scale=0.45]{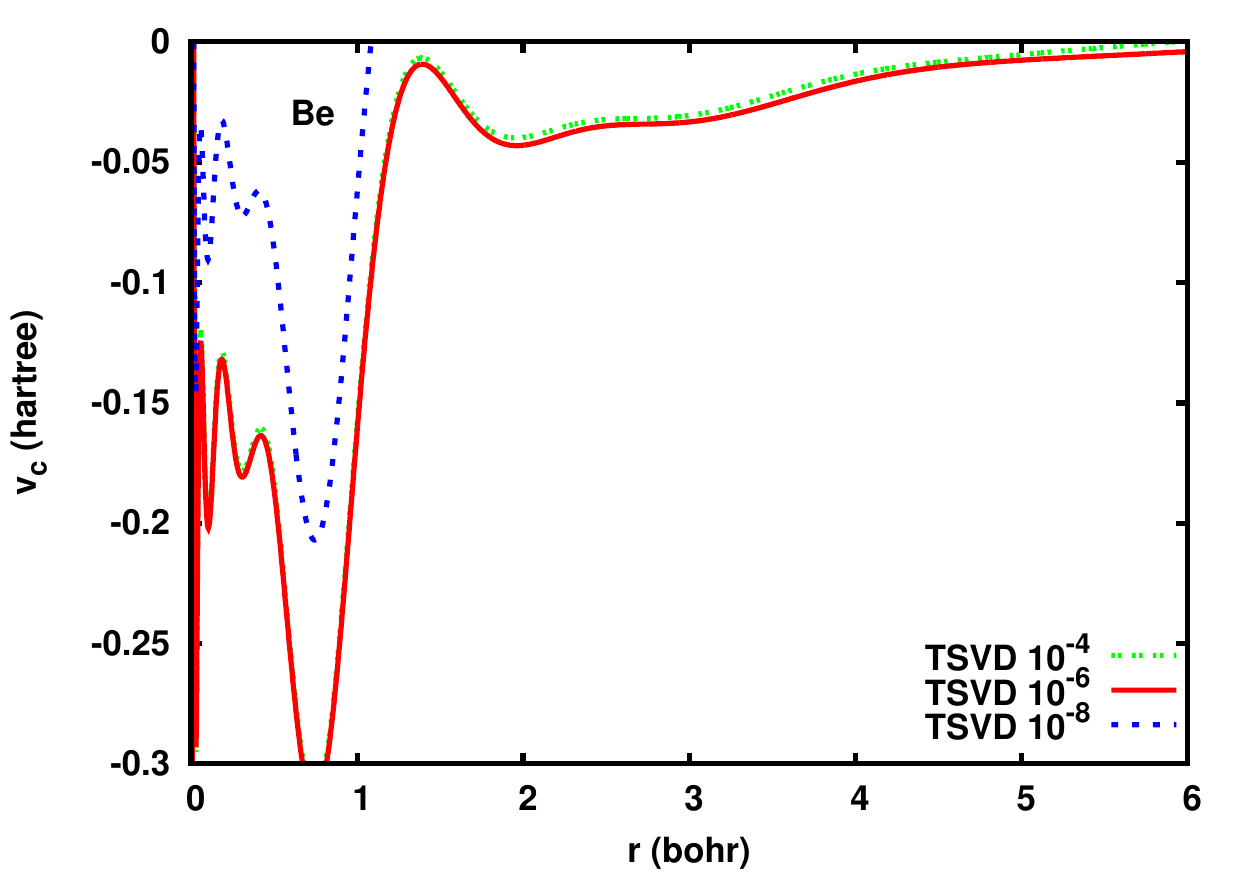}

\includegraphics[scale=0.45]{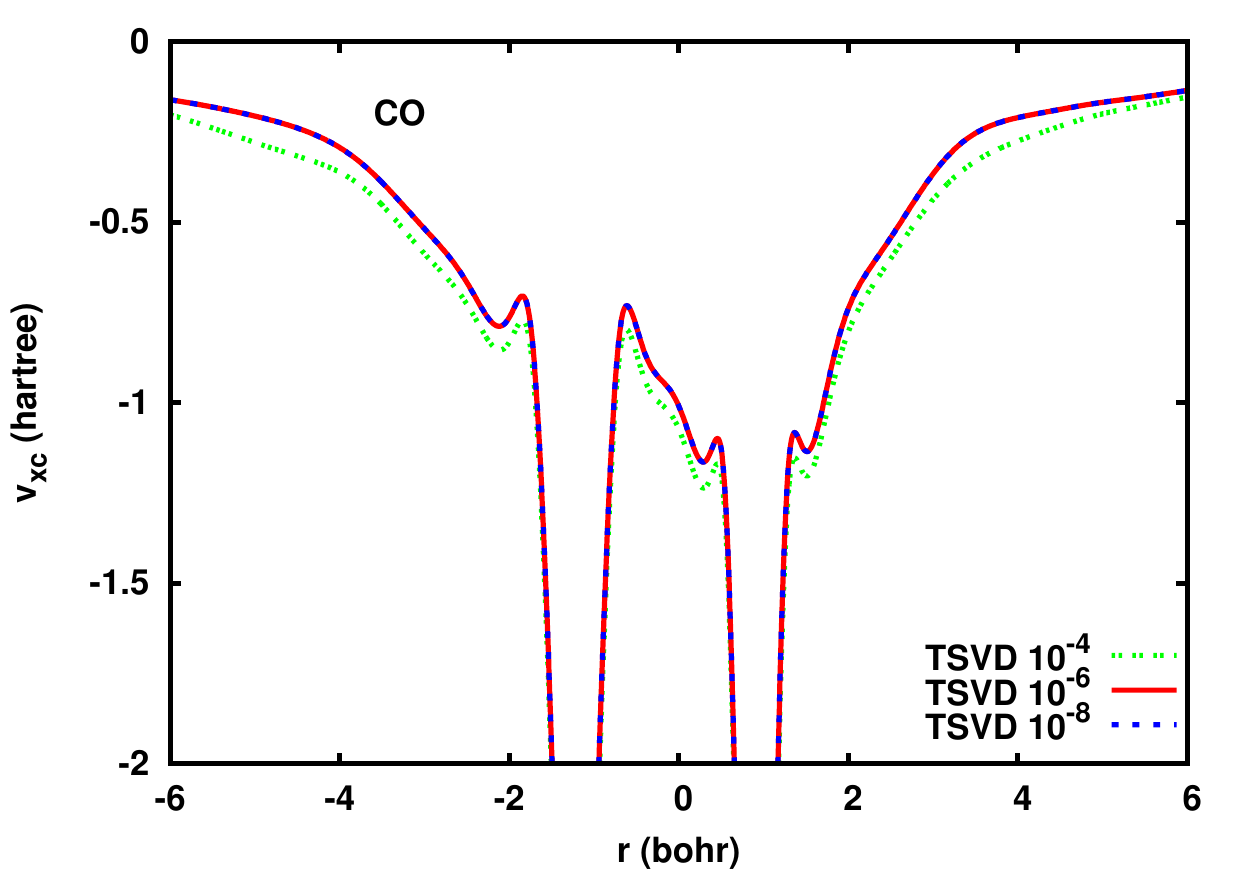}
\includegraphics[scale=0.45]{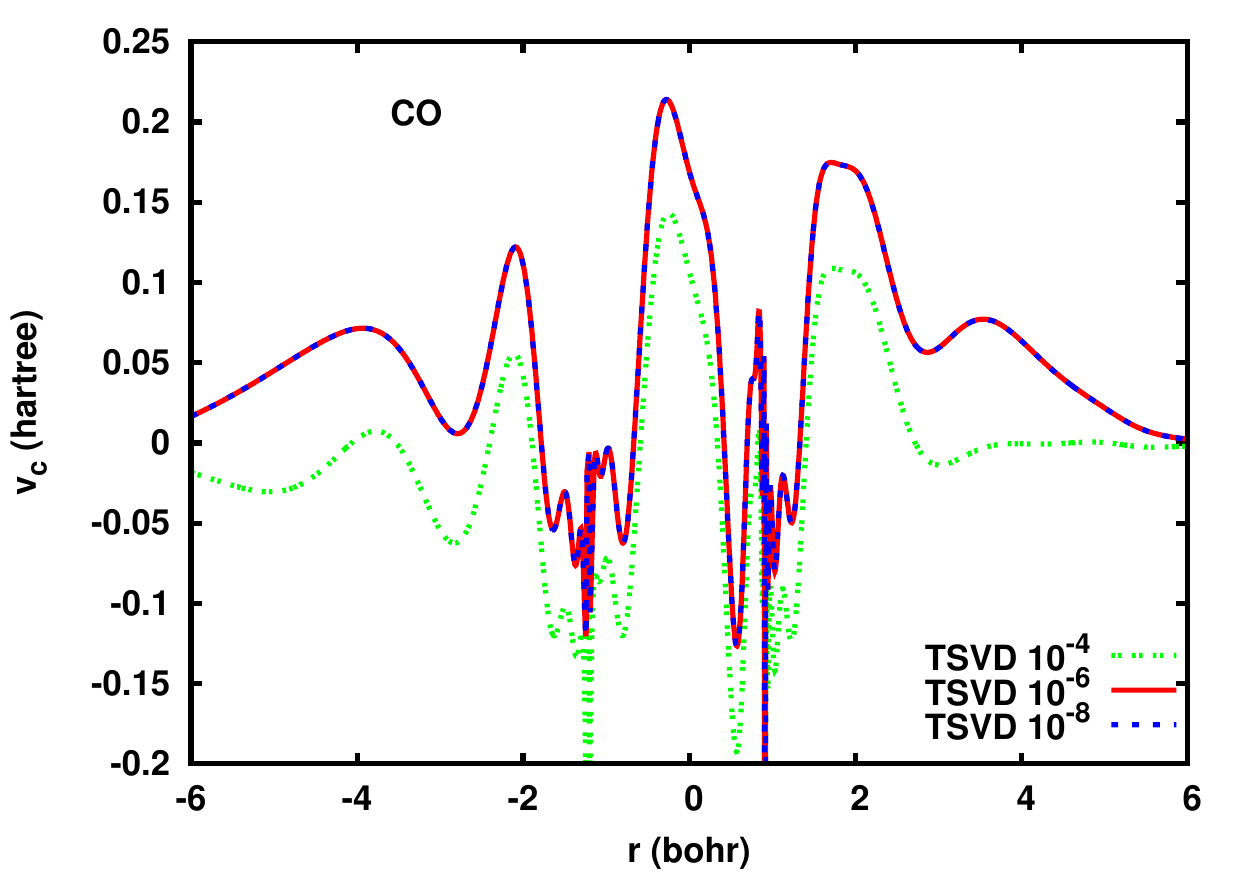}
\caption{Exchange-correlation and correlation potentials calculated with the OEP-1DH approximation using the BLYP functional at the recommended value $\lambda=0.65$ for the Be atom and the CO molecule using different TSVD cutoffs $10^{-4}$, $10^{-6}$, $10^{-8}$ for the pseudo-inversion of the linear-response function. For Be, the potentials for $10^{-4}$ and $10^{-6}$ are superimposed. For CO, the potentials for $10^{-6}$ and $10^{-8}$ are superimposed.}
\label{fig:svdcutoff}
\end{figure*}

The 1DH calculations have been performed with a development version of {\tt MOLPRO 2015}~\cite{Molproshort-PROG-15}, and the OEP-1DH ones with a development version of {\tt ACES II}~\cite{acesII}. In all calculations, we have used the B exchange~\cite{Bec-PRA-88} and the LYP correlation~\cite{LeeYanPar-PRB-88} density functionals, for $E_\text{x}^\text{DFA}$ and $E_\text{c}^\text{DFA}$, respectively. This choice was motivated by the fact that 1DH-BLYP was found to be among the one-parameter double-hybrid approximations giving the most accurate thermochemistry properties on average~\cite{ShaTouSav-JCP-11,TouShaBreAda-JCP-11,SouShaTou-JCP-14}. We expect however that the effect of the OEP self consistency to be similar when using other density functional approximations. The performance of both DH methods has been tested against a few atomic (He, Be, Ne, and Ar) and molecular (CO and H$_2$O) systems. For the latter, we considered the following equilibrium geometries: for CO $d(\text{C--O})=1.128 \text{\AA}$, and for  H$_2$O $d(\text{H--O})=0.959 \text{\AA}$ and $a(\text{H--O--H}) = 103.9^{\circ}$. In all cases, core excitations were included in the second-order correlation term.

In our OEP calculations, for convenience of implementation, the same basis set is used for expanding both the orbitals and the exchange-correlation potential. To ensure that the basis sets chosen were flexible enough for representation of orbitals and exchange-correlation potentials, all basis sets were constructed by full uncontraction of basis sets originally developed for correlated calculations, as in Refs.~\onlinecite{GraTeaSmiBar-JCP-11,GraTeaFabSmiBukDel-MP-14}. In particular, we employed an even tempered 20s10p2d basis for He, and an uncontracted ROOS--ATZP basis~\cite{widm90} for Be and Ne. For Ar, we used a modified basis set \cite{sos_oep2:prb} which combines s and p basis functions from the uncontracted ROOS--ATZP~\cite{widm90} with d and f functions coming from the uncontracted aug--cc--pwCVQZ basis set~\cite{peterson02}. In the case of both molecular systems, the uncontracted cc--pVTZ basis set of Dunning~\cite{dunning:1989:bas} was employed. For all OEP calculations standard convergence criteria were enforced, corresponding to maximum deviations in density-matrix elements of 10$^{-8}$. In practice, the use of the same basis set for expanding both the orbitals and the exchange-correlation potential leads to the necessity of truncating the auxiliary function space by the TSVD method for constructing the pseudo-inverse of the linear-response function. The convergence of the potentials with respect to the TSVD cutoff was studied. Figure~\ref{fig:svdcutoff} shows the example of the convergence of the exchange-correlation and correlation potentials of the Be atom and the CO molecule. For Be, the potentials obtained with the 10$^{-4}$ and 10$^{-6}$ cutoffs are essentially identical, while for the 10$^{-8}$ cutoff the exchange-correlation potential has non-physical oscillations and the correlation potential diverges. For CO, the potentials obtained with the 10$^{-4}$ cutoff are significantly different from the potentials obtained with the 10$^{-6}$ cutoff, while no difference can be seen between the potentials obtained with the 10$^{-6}$ and 10$^{-8}$ cutoffs. A cutoff of 10$^{-6}$ was thus chosen for all systems to achieve a compromise between convergence and numerical stability.

In order to assess the quality of the results obtained with the standard and OEP-based DH methods, we considered several reference data. We used estimated exact total energies extracted from numerical calculations~\cite{ChaGwaDavParFro-PRA-93} for He, Be, Ne, Ar, and from quadratic configuration-interaction calculations extrapolated to the complete basis set~\cite{KloBakJorOlsHel-JPB-99} for CO and H$_2$O. We also used reference data from coupled-cluster singles-doubles with perturbative triples [CCSD(T)]~\cite{purvis82,pople87,scuseria88,Raghavachari1989479} calculations performed with the same basis sets. In particular, these CCSD(T) calculations yielded densities which were used as input for generating reference KS potentials by inversion of the KS equations~\cite{ColSav-JCP-99,wu:2003:wy}, using the computational setup described in Refs.~\onlinecite{GraTeaSmiBar-JCP-11,GraTeaFabSmiBukDel-MP-14}. We also used estimated exact total energies extracted from numerical calculations~\cite{ChaGwaDavParFro-PRA-93} for He, Be, Ne, Ar, and from quadratic configuration-interaction calculations extrapolated to the complete basis set~\cite{KloBakJorOlsHel-JPB-99} for CO and H$_2$O.

\section{Results and discussion}
\label{sec:results}

\subsection{Total energies}

\begin{figure*}
\includegraphics[scale=0.45]{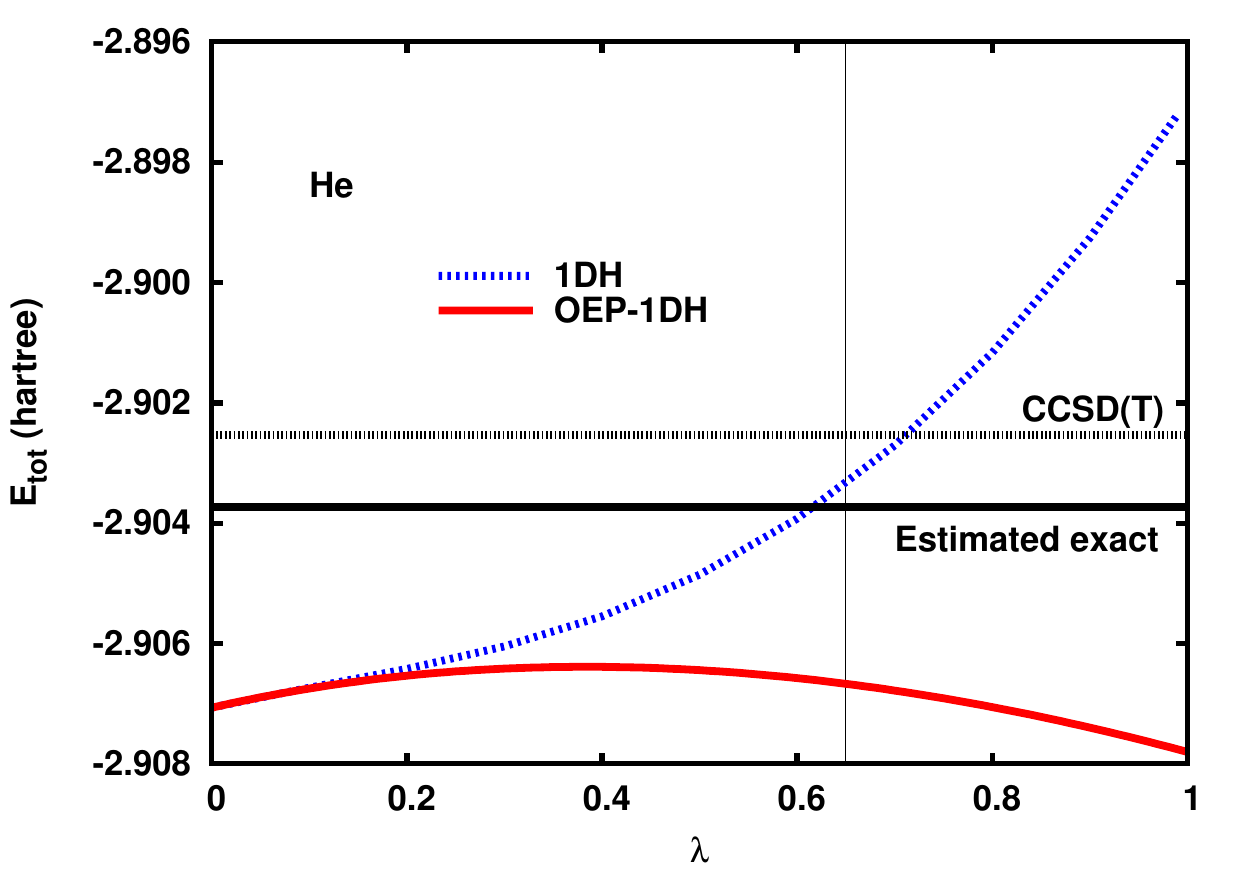}
\includegraphics[scale=0.45]{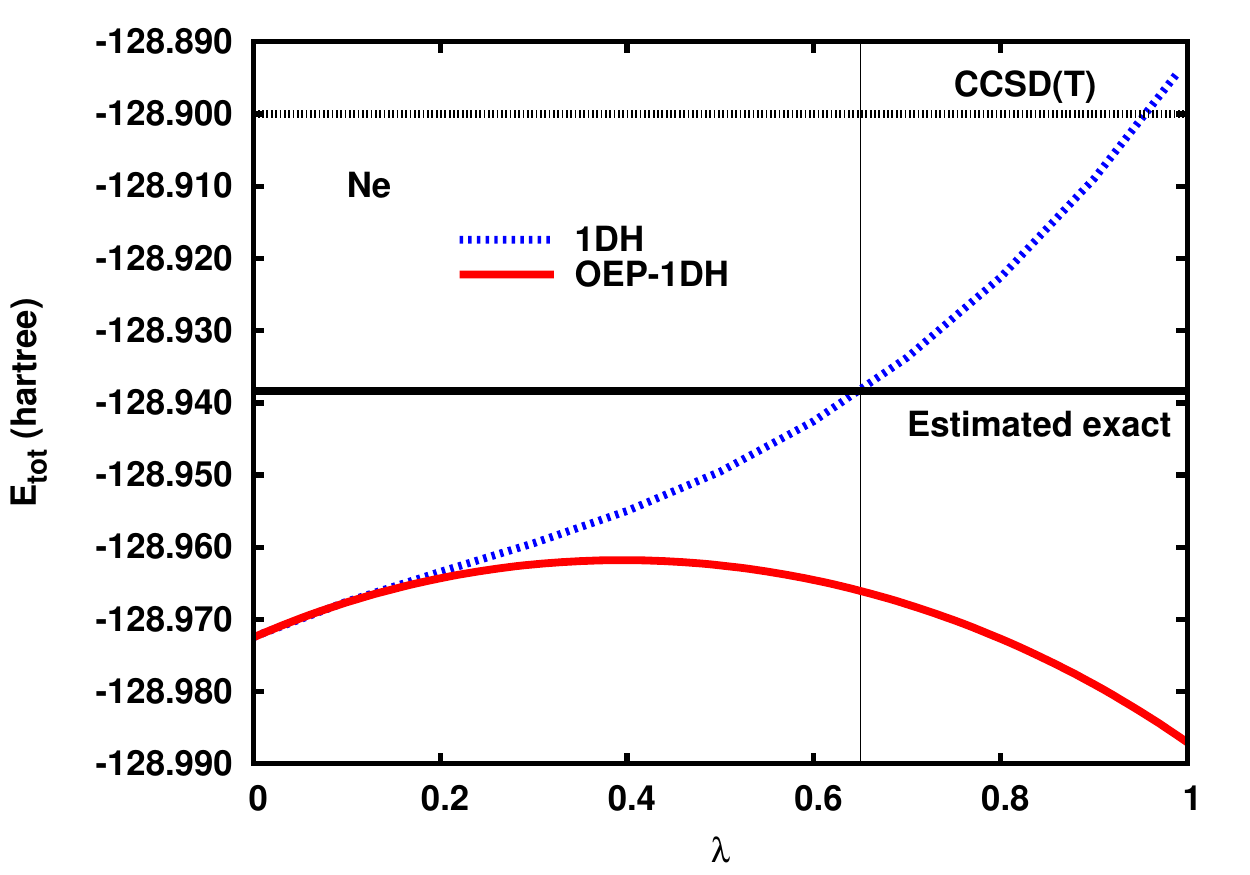}
\includegraphics[scale=0.45]{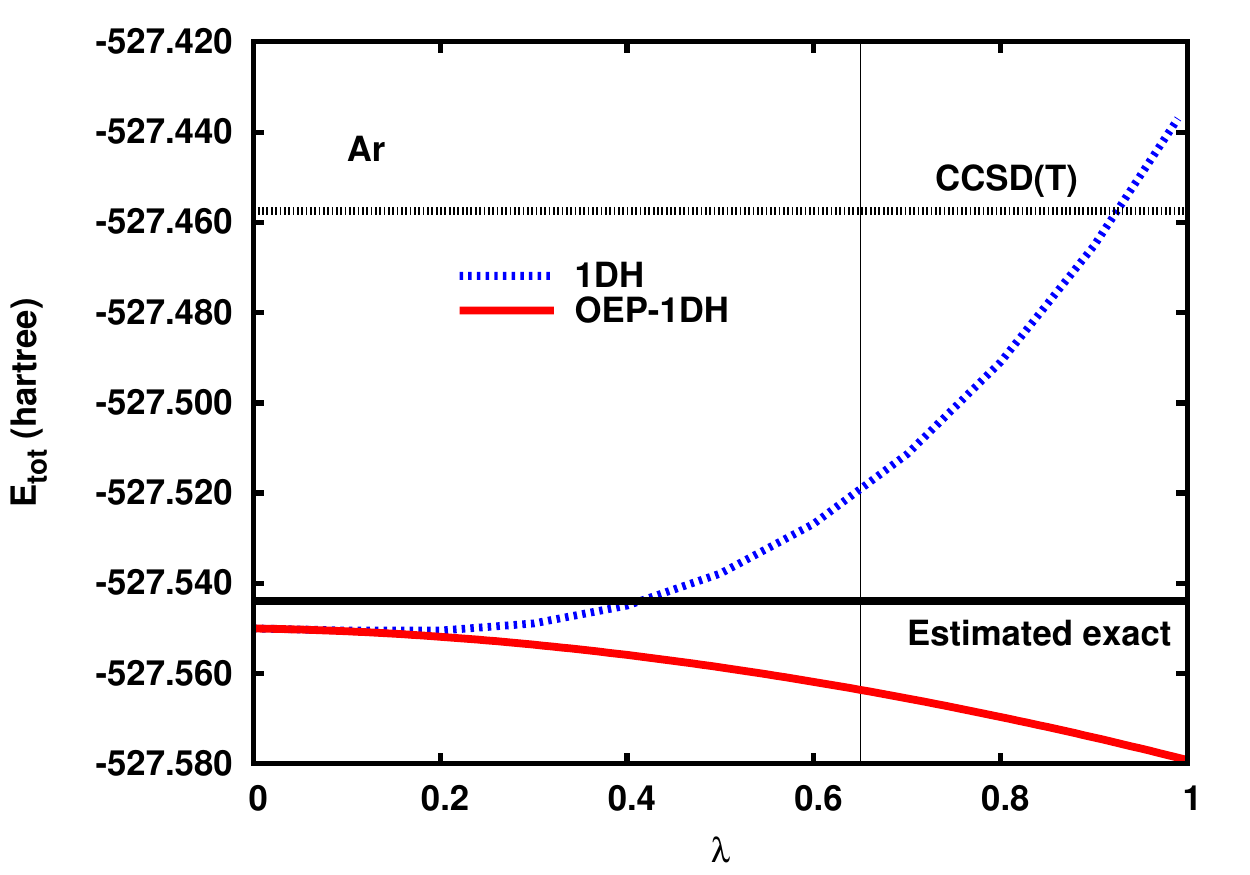}
\includegraphics[scale=0.45]{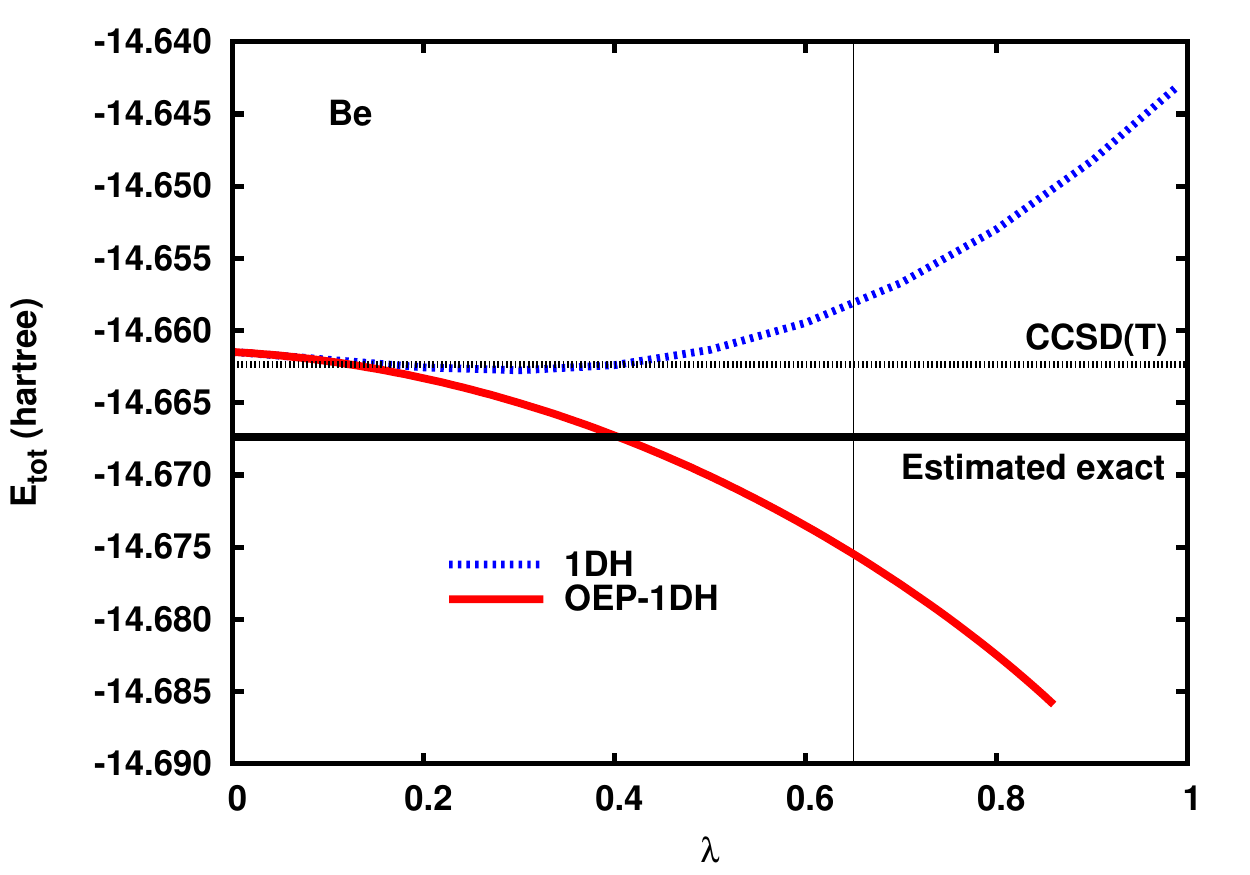}
\includegraphics[scale=0.45]{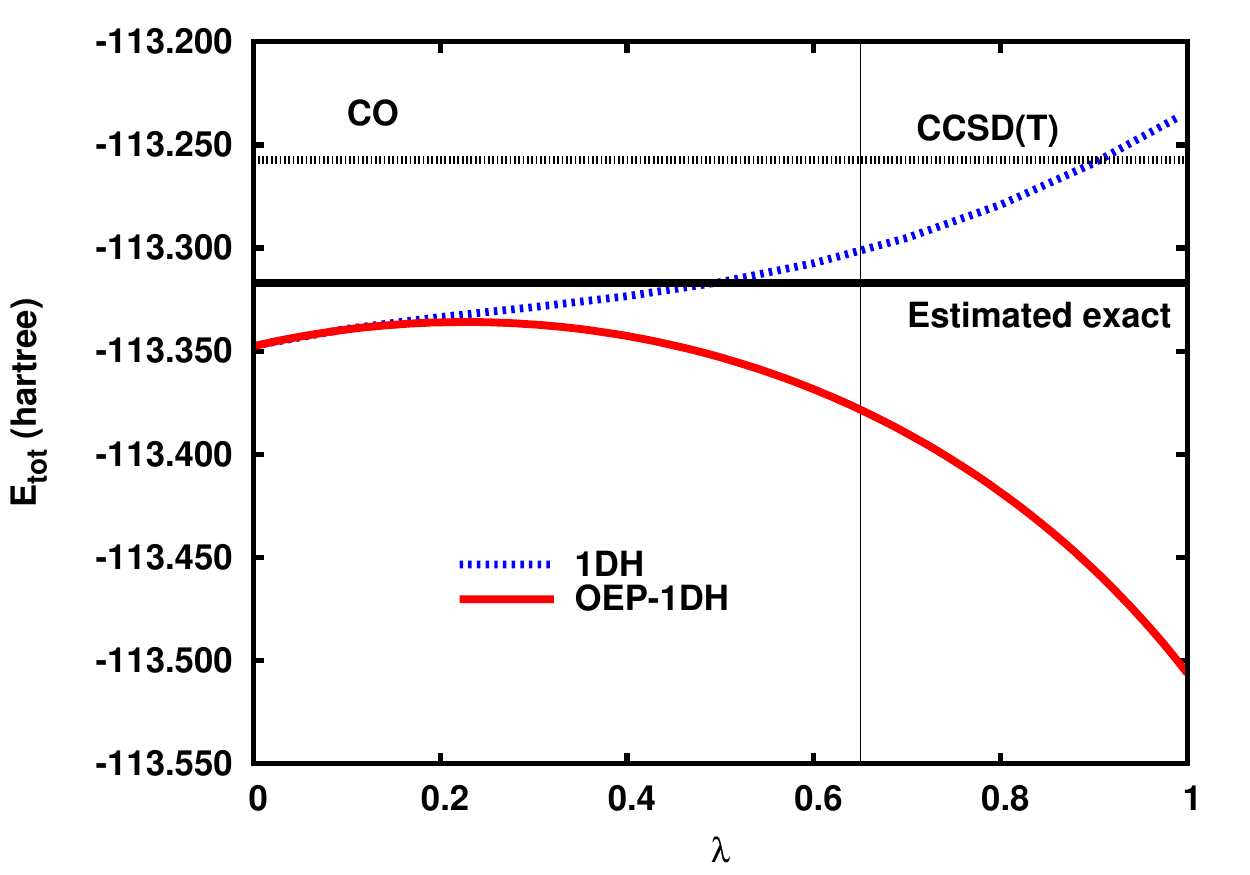}
\includegraphics[scale=0.45]{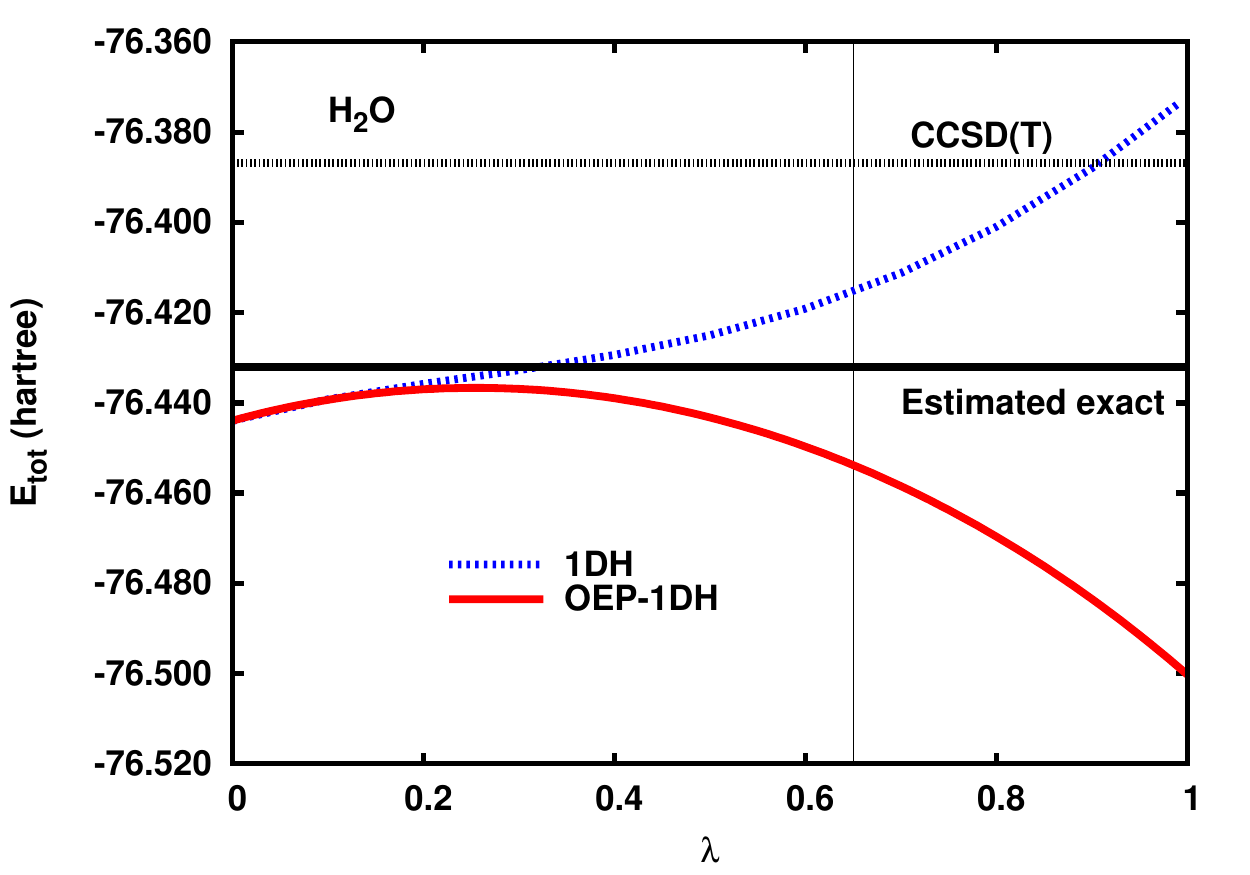}
\caption{Total energies calculated with the 1DH and OEP-1DH approximations with the BLYP functional as a function of $\lambda$. As reference values, CCSD(T) total energies calculated with the same basis sets are given, as well as estimated exact total energies taken from Ref.~\onlinecite{ChaGwaDavParFro-PRA-93} for He, Be, Ne, Ar, and from Ref.~\onlinecite{KloBakJorOlsHel-JPB-99} for CO and H$_2$O. The vertical lines correspond to $\lambda=0.65$, i.e. the value recommended for 1DH with the BLYP functional in Ref.~\onlinecite{ShaTouSav-JCP-11}. For Be, the OEP-1DH calculations are unstable for $\lambda >0.86$.
}
\label{fig:etot}
\end{figure*}

Figure~\ref{fig:etot} shows the total energy of each system as a function of $\lambda$ calculated with the 1DH and OEP-1DH approximations. For comparison, CCSD(T) total energies calculated with the same basis sets and estimated exact total energies taken from Refs.~\onlinecite{ChaGwaDavParFro-PRA-93,KloBakJorOlsHel-JPB-99} are also reported. Note that the CCSD(T) total energies are significantly higher than the estimated exact energies, which is mostly due to the incompleteness of the basis sets used. Since the explicit density-functional contribution of the DH calculations does not suffer from this large basis incompleteness error, we prefer to use as reference the estimated exact energies.

At $\lambda=0$, both DH methods reduce to standard KS using the BLYP functional, which tends to overestimate the total energy by about 3 to 30 mhartree, except for Be atom where it is underestimated (which may be connected to the presence of an important static correlation contribution in this system). 
At $\lambda=1$, the 1DH approximation reduces to standard MP2 (with HF orbitals), while the OEP-1DH approximation reduces to OEP-GL2 (i.e., the same MP2 total energy expression but with fully self-consistently optimized OEP orbitals)~\cite{GraHirIvaBar-JCP-02,BarGraHirIva-JCP-05}. 
Standard MP2 systematically underestimates the total energy on magnitude (up to more than 100 mhartree for Ar) which is partly due to the missing correlation contribution beyond second order and to the incompleteness of the basis sets used. On the opposite, OEP-GL2 systematically gives too negative total energies, as already known~\cite{BarGraHirIva-JCP-05,grabowski:2007:ccpt2}. For example, for CO the OEP-GL2 total energy is more than 150 mhartree too low. Note that for Be the OEP-GL2 calculation is unstable, as already reported~\cite{engel:2005:oeppt2,BarGraHirIva-JCP-05,MorWuYan-JCP-05}. The fact that OEP-GL2 gives much more negative total energies than standard MP2 should be connected to the fact that the HOMO-LUMO orbital energy gap is much smaller with OEP-GL2 orbitals than with HF orbitals (see results in Sections~\ref{sec:ip} and~\ref{sec:ea}).

In between the extreme values $\lambda=0$ and $\lambda=1$, the 1DH and OEP-1DH approximation give smooth total energy curves, which start to visually differ for $\lambda \gtrsim 0.2$. Note that for Be the OEP-1DH calculations are stable for $\lambda\leq0.86$. At $\lambda=0.65$, which is the value recommended in Ref.~\onlinecite{ShaTouSav-JCP-11} for 1DH with the BLYP functional, both 1DH and OEP-1DH give more accurate total energies than their respective $\lambda=1$ limits (i.e., MP2 and OEP-GL2), but do not perform necessarily better than the $\lambda=0$ limit (KS BLYP). Depending on the system considered, at $\lambda=0.65$, the 1DH total energy is either more accurate or about equally accurate than the OEP-1DH total energy. Thus, even though OEP-1DH provides an important improvement over OEP-GL2, we conclude that the self-consistent optimization of the orbitals in the 1DH approximation (i.e., going from 1DH to OEP-1DH) does not lead to improved ground-state total energies for the few systems considered here. We expect that a similar conclusion generally holds for ground-state energy differences such as atomization energies, similarly to what has been found for the case of the hybrid approximations~\cite{Kar-JCP-03}.

\subsection{HOMO orbital energies and ionization potentials}
\label{sec:ip}

\begin{figure*}
\includegraphics[scale=0.45]{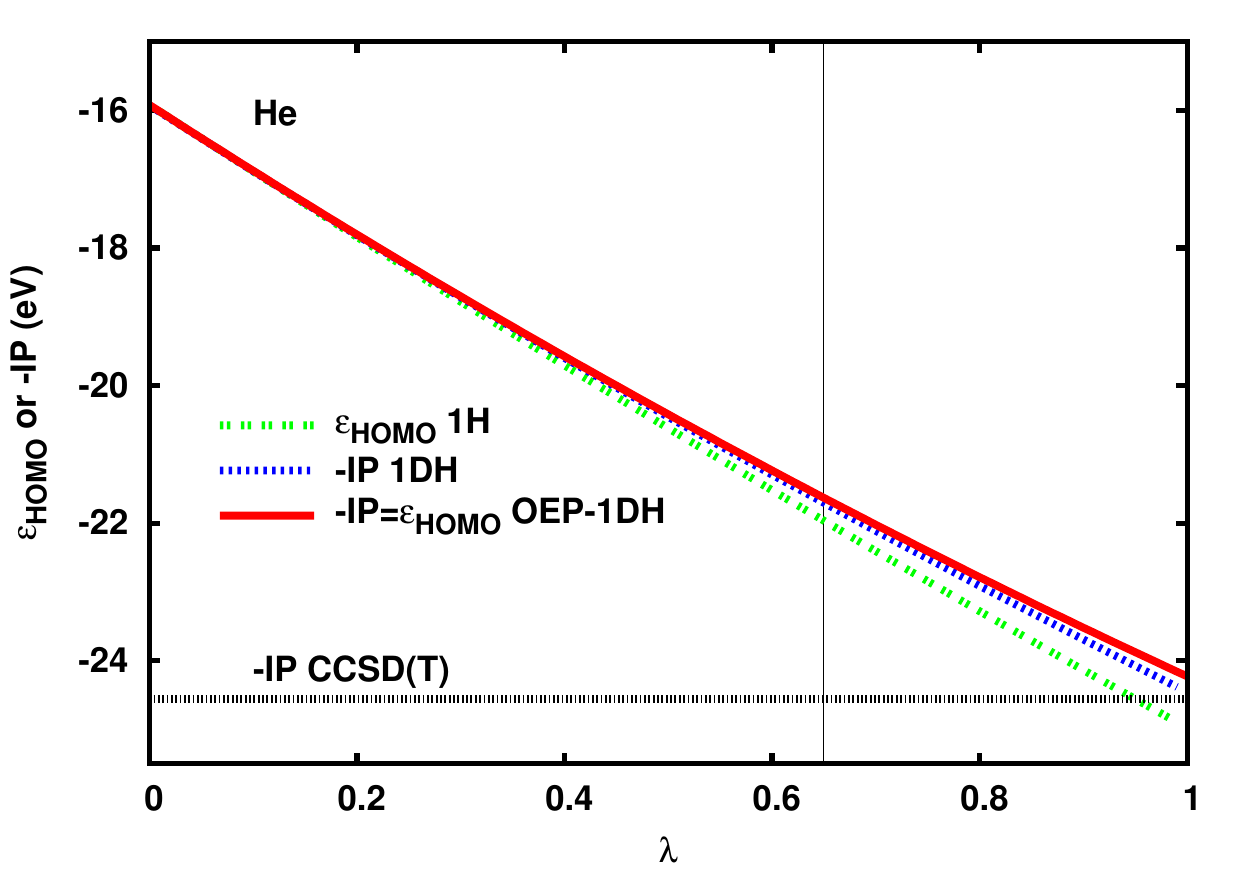}
\includegraphics[scale=0.45]{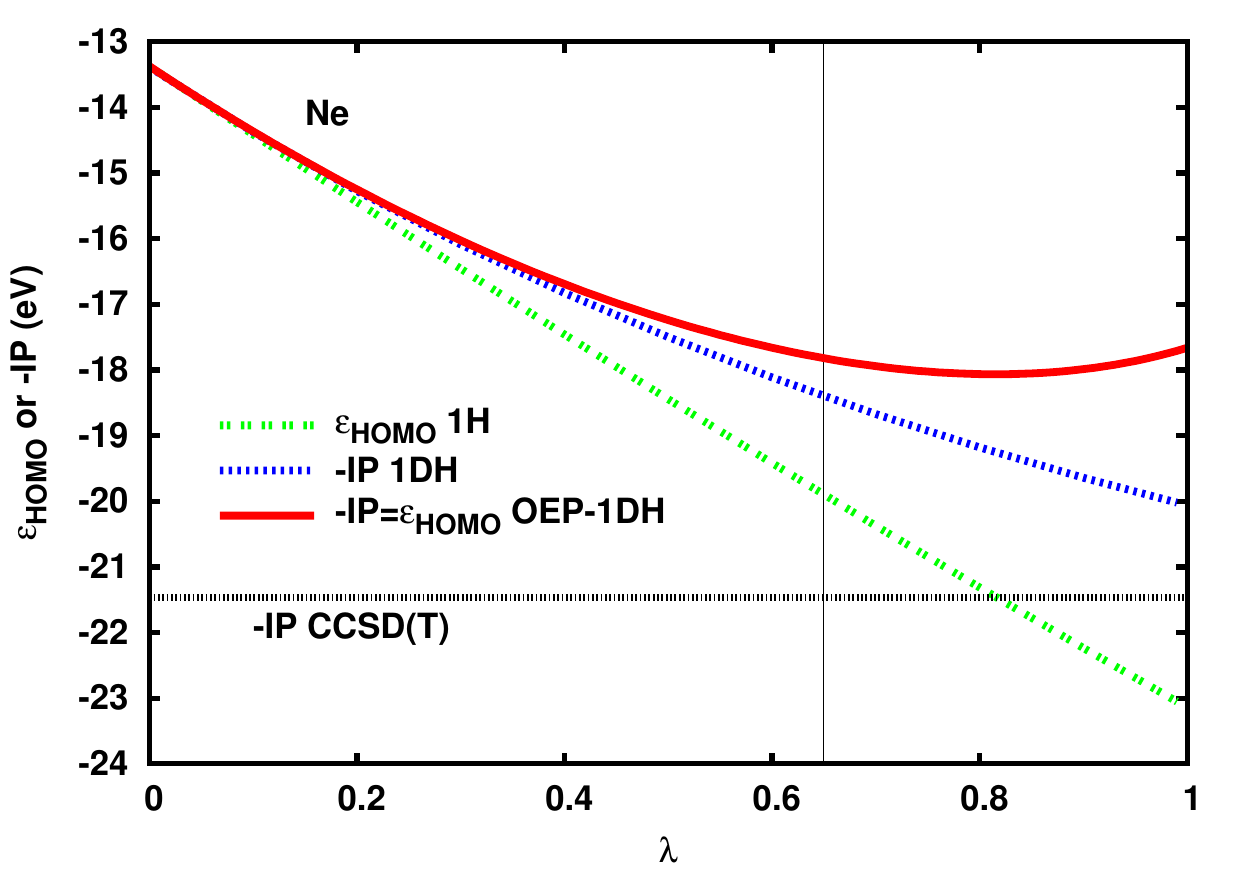}
\includegraphics[scale=0.45]{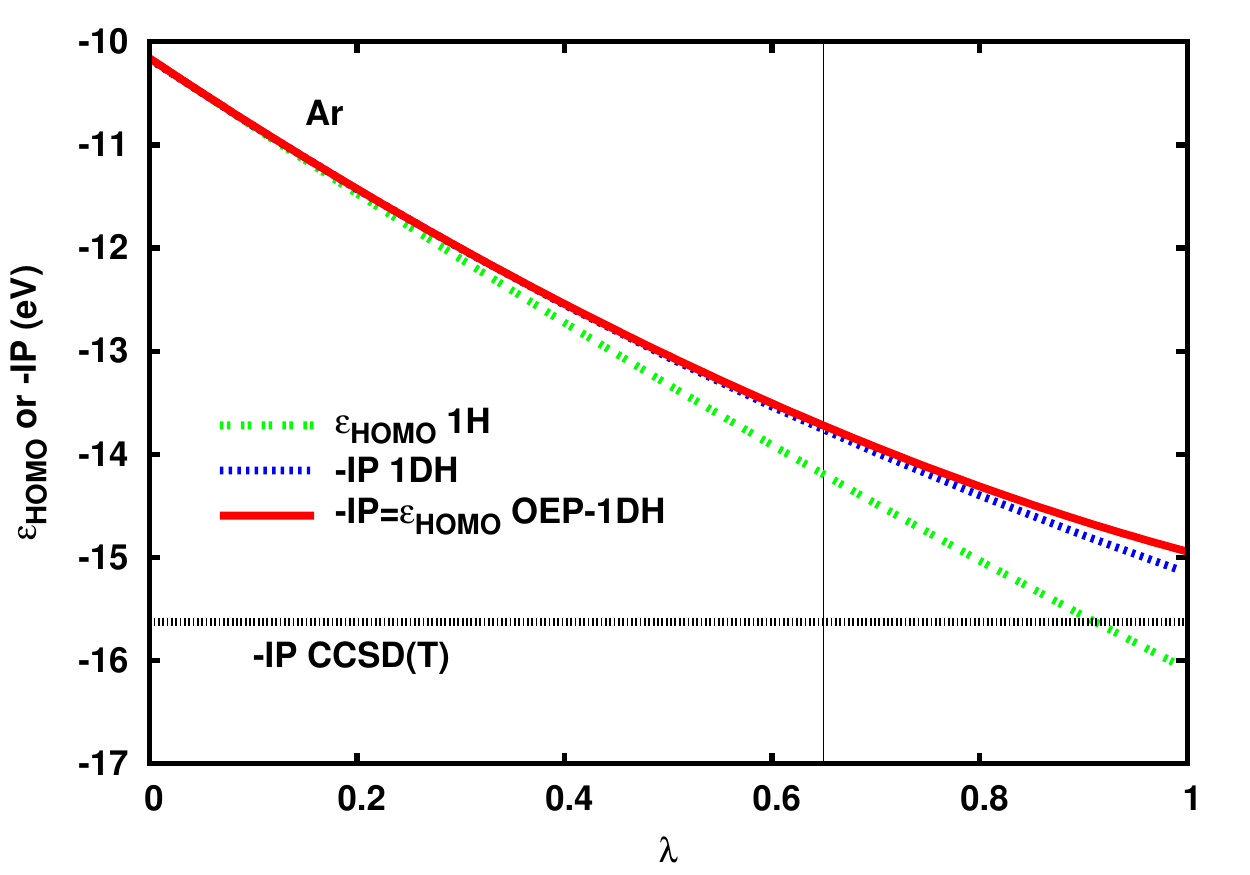}
\includegraphics[scale=0.45]{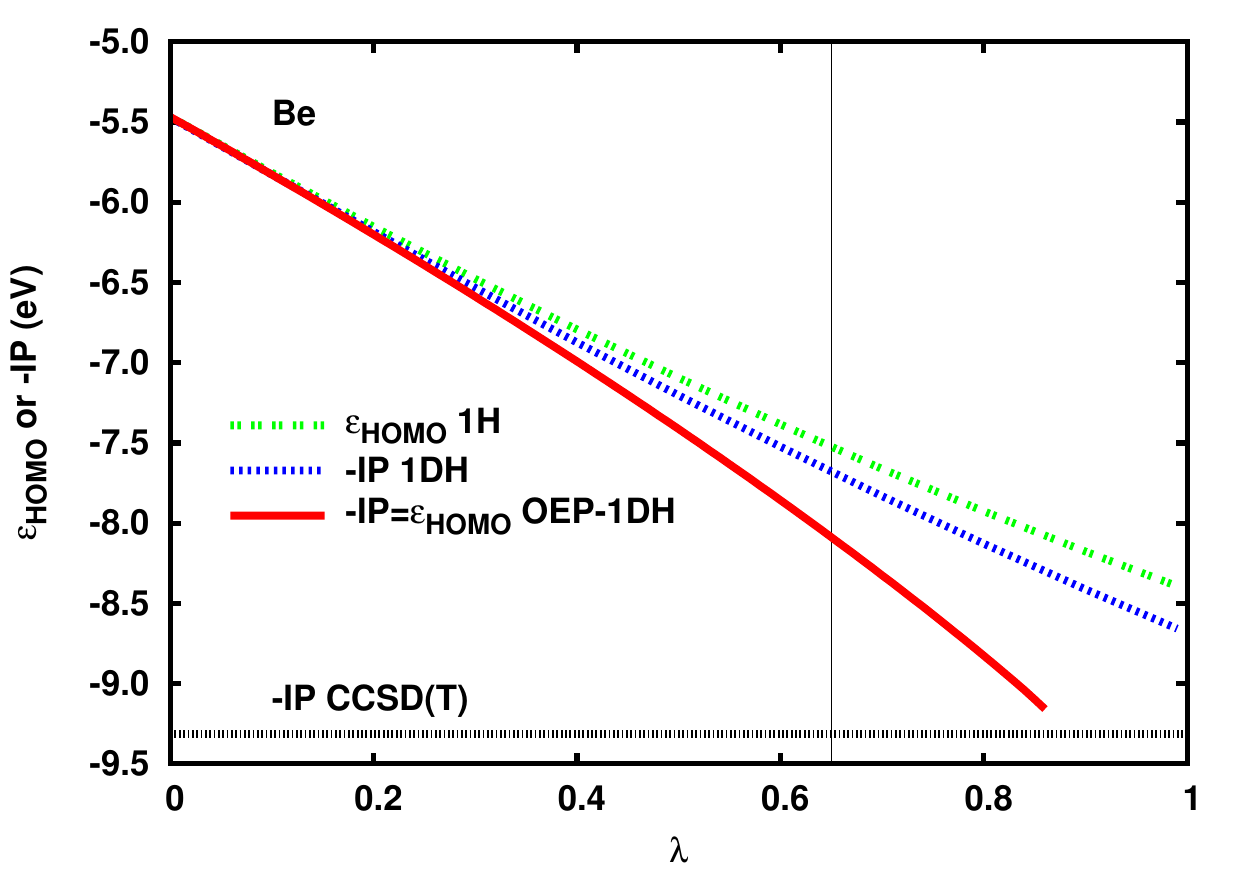}
\includegraphics[scale=0.45]{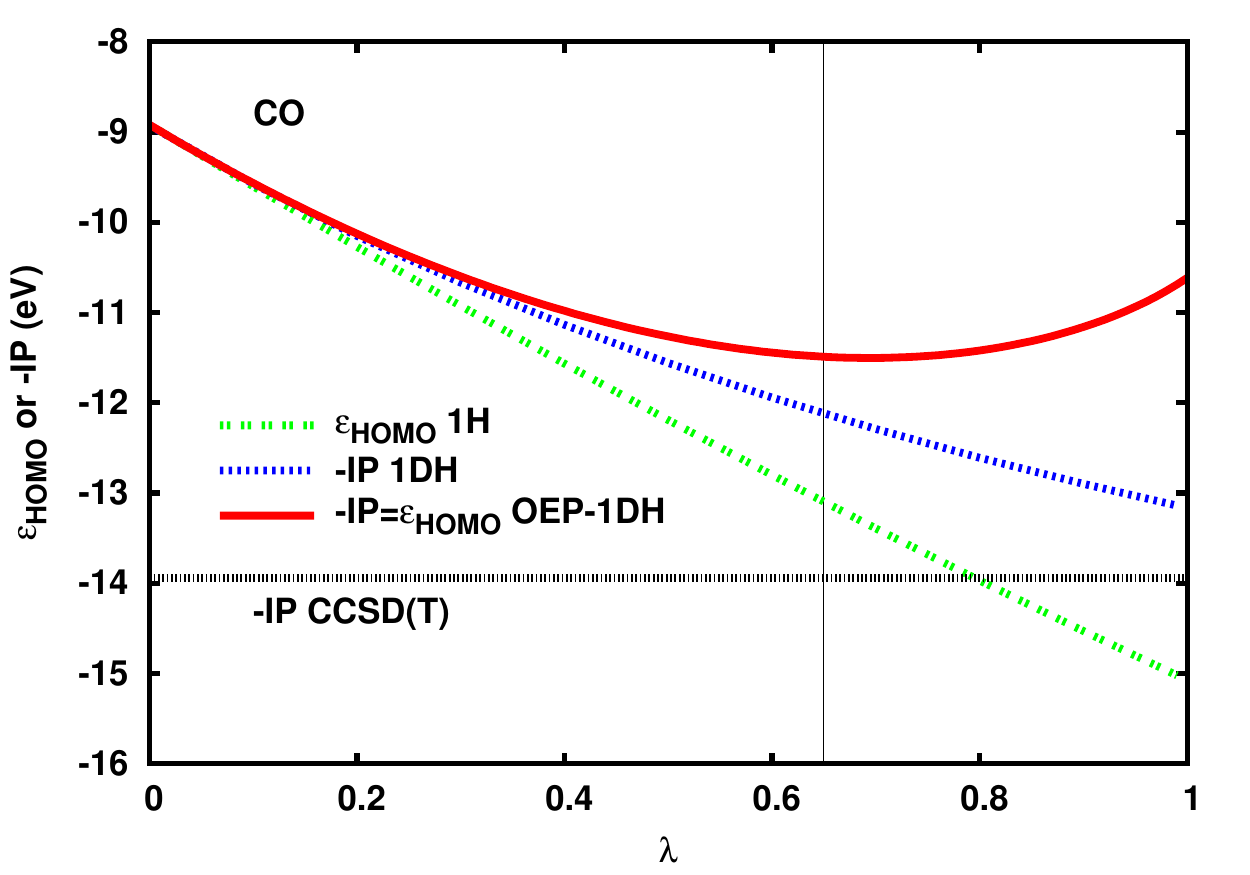}
\includegraphics[scale=0.45]{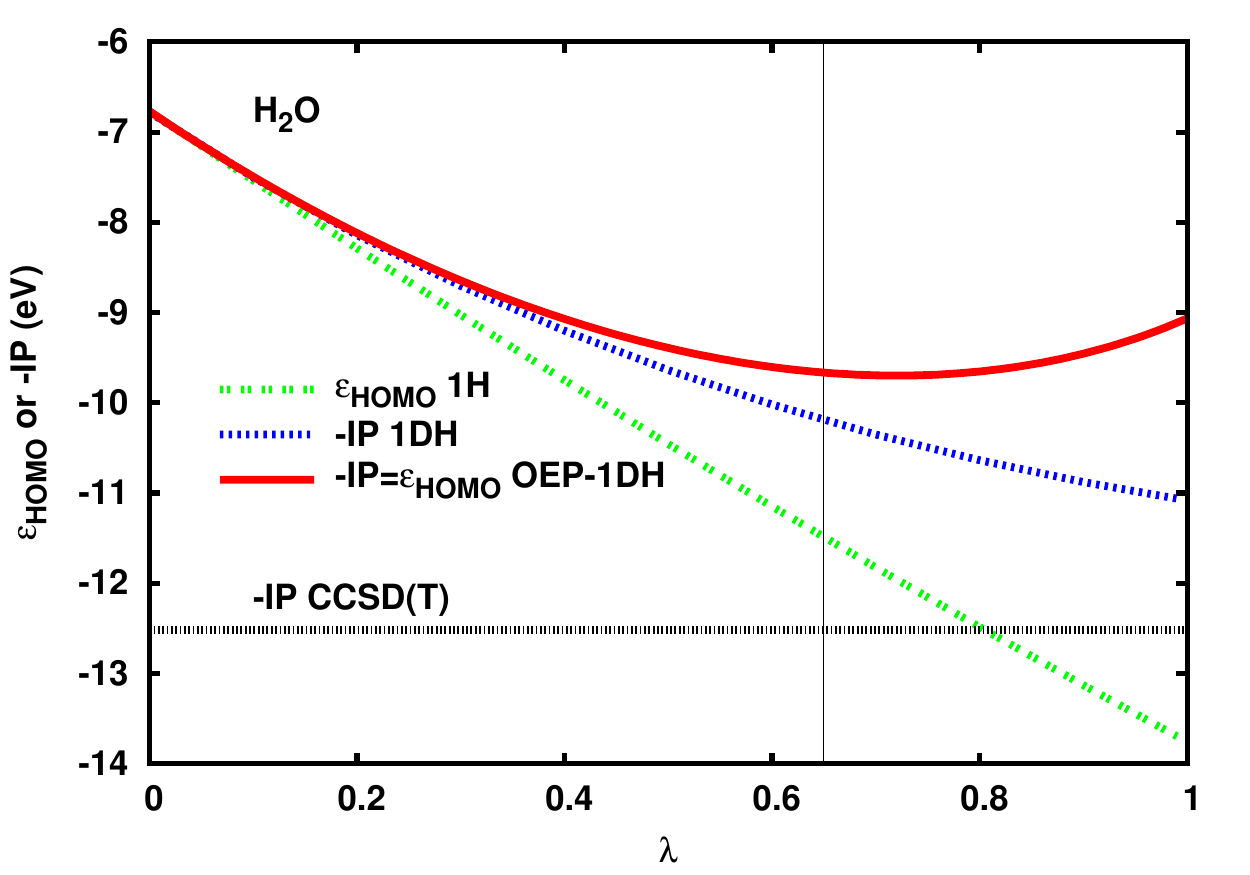}
\caption{HOMO orbital energies in the 1H approximation [Eq.~(\ref{hnlpsi})] and minus IPs in the 1DH [Eq.~(\ref{IP1DH})] and OEP-1DH [Eq.~(\ref{IPOEP1DH})] approximations using the BLYP functional as a function of $\lambda$. The reference values were calculated as CCSD(T) total energy differences with the same basis sets. The vertical lines correspond to $\lambda=0.65$, i.e. the value recommended for 1DH with the BLYP functional in Ref.~\onlinecite{ShaTouSav-JCP-11}. For Be, the OEP-1DH calculations are unstable for $\lambda >0.86$.
}
\label{fig:ip}
\end{figure*}

Figure~\ref{fig:ip} reports, for each system, the HOMO orbital energy in the 1H approximation [Eq.~(\ref{hnlpsi})] and minus the IPs in the 1DH [Eq.~(\ref{IP1DH})] and OEP-1DH [Eq.~(\ref{IPOEP1DH})] approximations as a function of $\lambda$. The reference IPs are from CCSD(T) calculations with the same basis sets. The HOMO orbital energy in the 1H approximation represents the simplest approximation to $-$IP available when doing a 1DH calculation. At $\lambda=0$, the 1H approximation reduces to standard KS with the BLYP functional, and we recover the well-known fact the HOMO orbital energy is much too high (underestimating the IP by about 4 to 8 eV for the systems considered here) with a semilocal DFA like BLYP, most likely due to the self-interaction error in the exchange density functional. At $\lambda=1$, the 1H approximation reduces to standard HF, and in this case the HOMO orbital energy is a much better estimate of $-$IP, overestimating the IP by about 1 eV except for Be where it is underestimated by about the same amount. In between $\lambda=0$ and $\lambda=1$, the 1H HOMO orbital energy varies nearly linearly with $\lambda$, which suggests that the $\lambda$-dependence is largely dominated by the exchange potential in Eq.~(\ref{vxc1H}). At $\lambda=0.65$, the 1H HOMO orbital energy is always higher than the reference $-$IP, by about 1 or 2 eV depending on the system considered.

The IPs obtained with the 1DH method, i.e. taking into account the MP2 correlation term, are smaller than the 1H IPs for all $\lambda$ and all systems considered here, with the exception of Be for which it is a bit larger. This is consistent with previous works which found that IPs calculating by taking the derivative of the MP2 total energy with respect the electron number are generally too small~\cite{CohMorYan-JCTC-09,SuXu-JCTC-15}. At $\lambda=0.65$, 1DH gives IPs that are underestimated by about 2 or 3 eV, which is similar to the average accuracy obtained with the two-parameter B2-PLYP double-hybrid approximation~\cite{SuYanMorXu-JPCA-14}. The effect of self-consistency, i.e. going from 1DH to OEP-1DH, is to further reduce the IPs, except for Be for which it increases it. For He the differences between 1DH and OEP-1DH are very small, which may not be surprising since for such a two-electron system the HF and EXX potentials have the same action on occupied orbitals. At $\lambda=1$, OEP-GL2 gives IPs which are generally not very accurate (see also Ref.~\onlinecite{SmiDelBukGraFab-JCC-16}). In particular, for Ne, CO, and H$_2$O, OEP-GL2 underestimates the IP by more than 3 eV. As a consequence, for these systems, for $\lambda \gtrsim 0.5$ self-consistency only deteriorates the accuracy of the IPs. We note that better IPs could be obtained using modified second-order correlated OEP approximations~\cite{SmiDelBukGraFab-JCC-16}.

\subsection{LUMO orbital energies and electronic affinities}
\label{sec:ea}

\begin{figure*}
\includegraphics[scale=0.45]{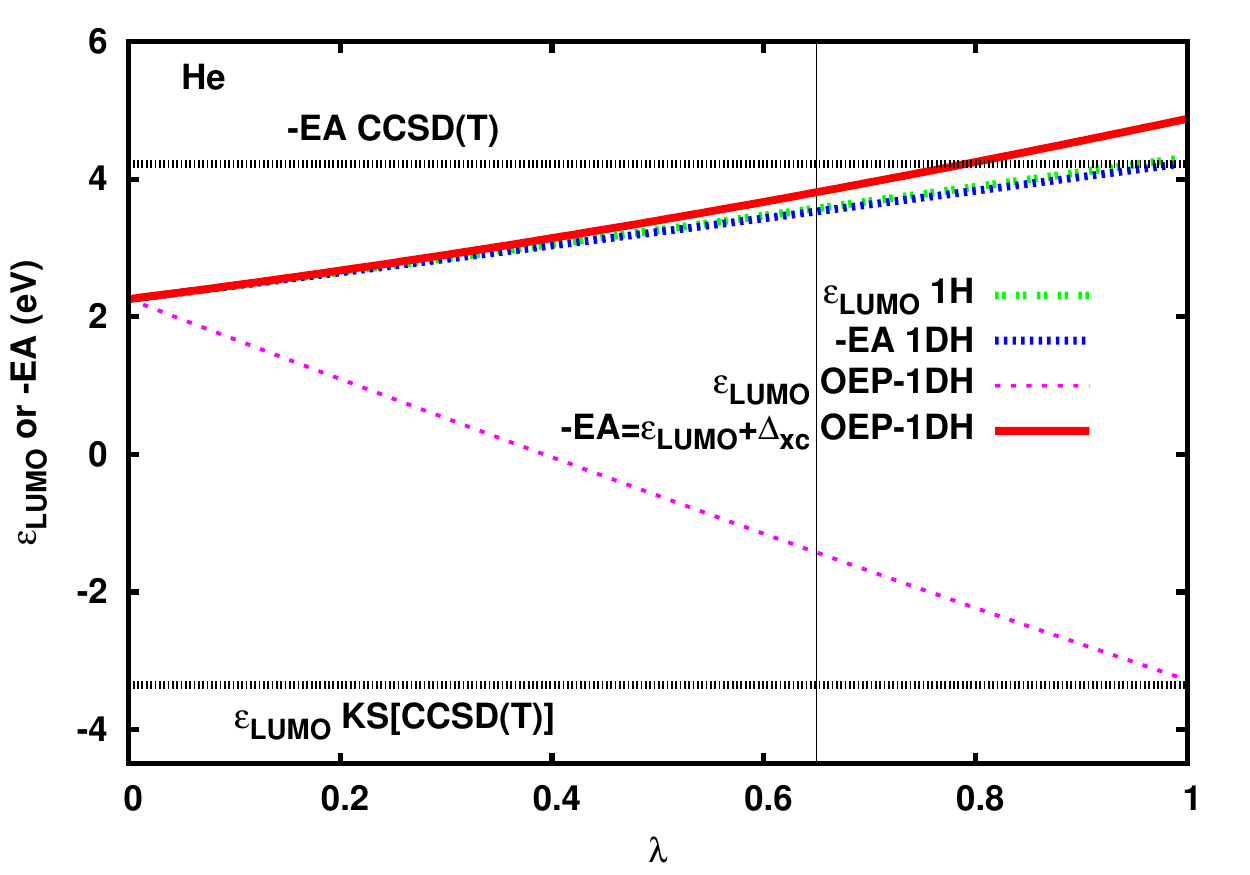}
\includegraphics[scale=0.45]{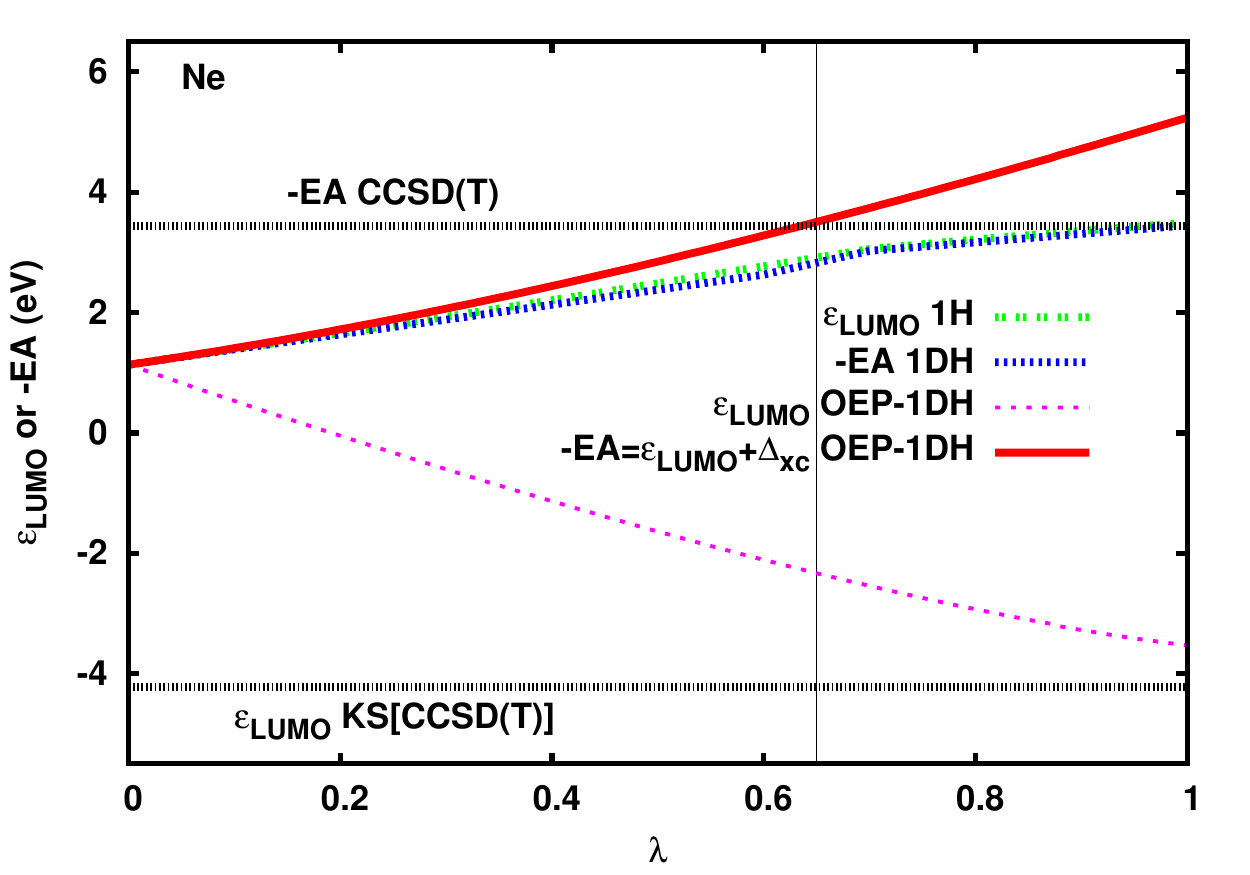}
\includegraphics[scale=0.45]{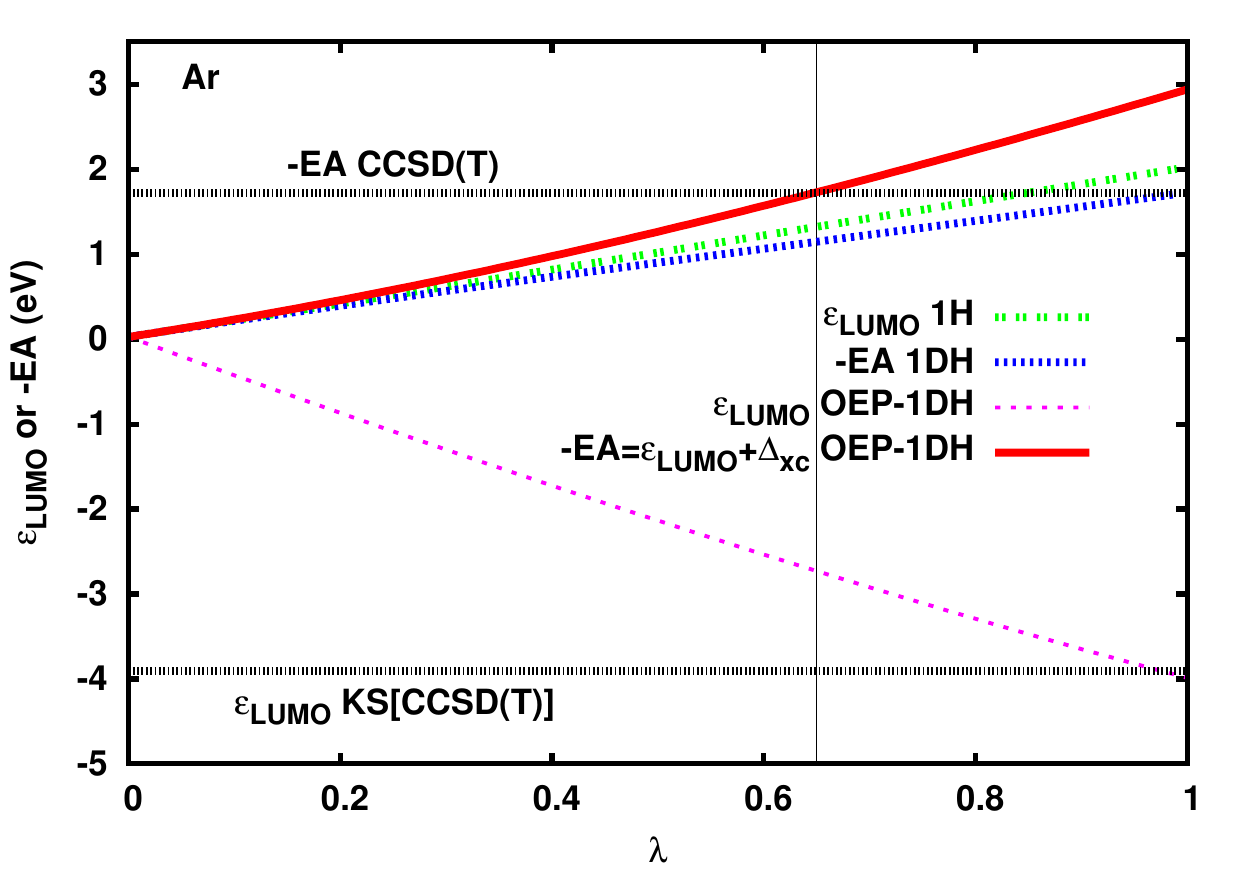}
\includegraphics[scale=0.45]{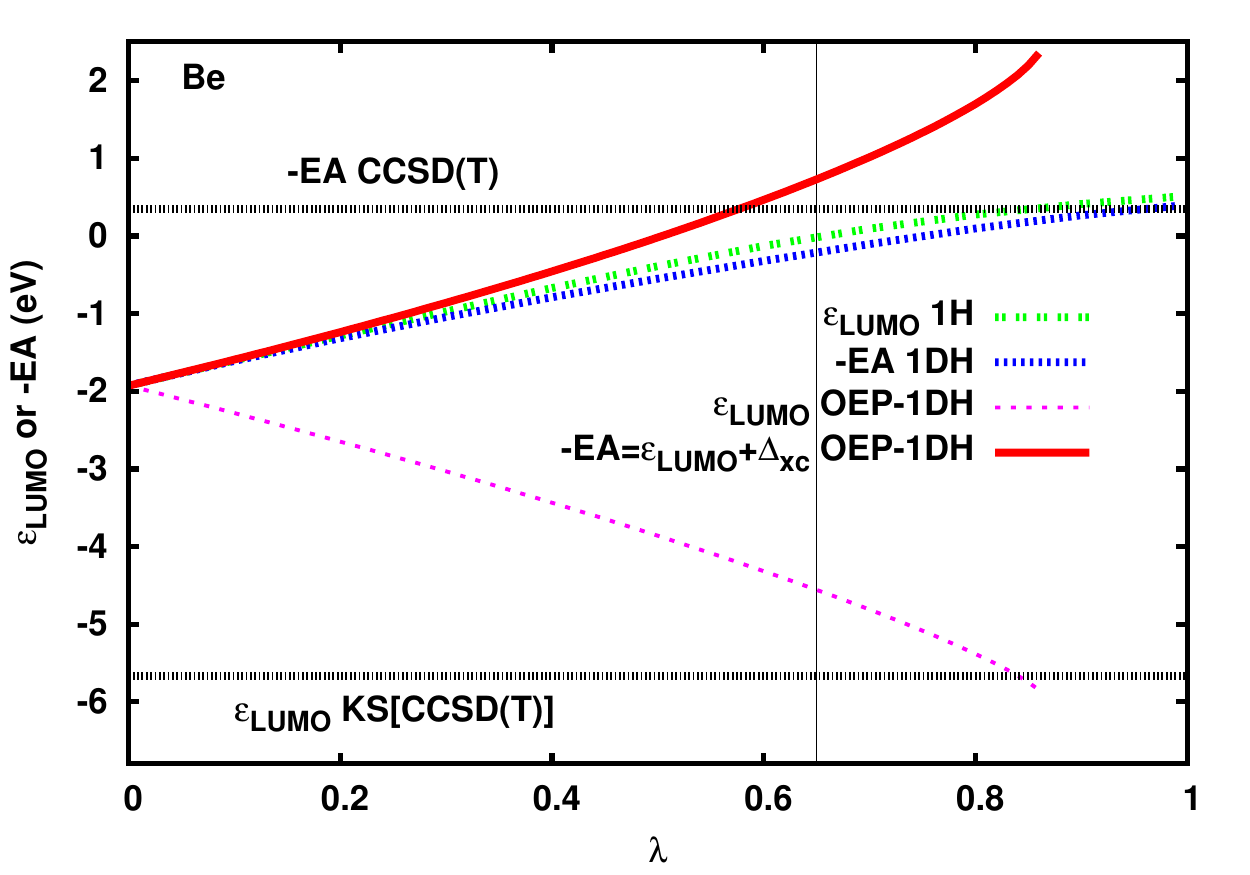}
\includegraphics[scale=0.45]{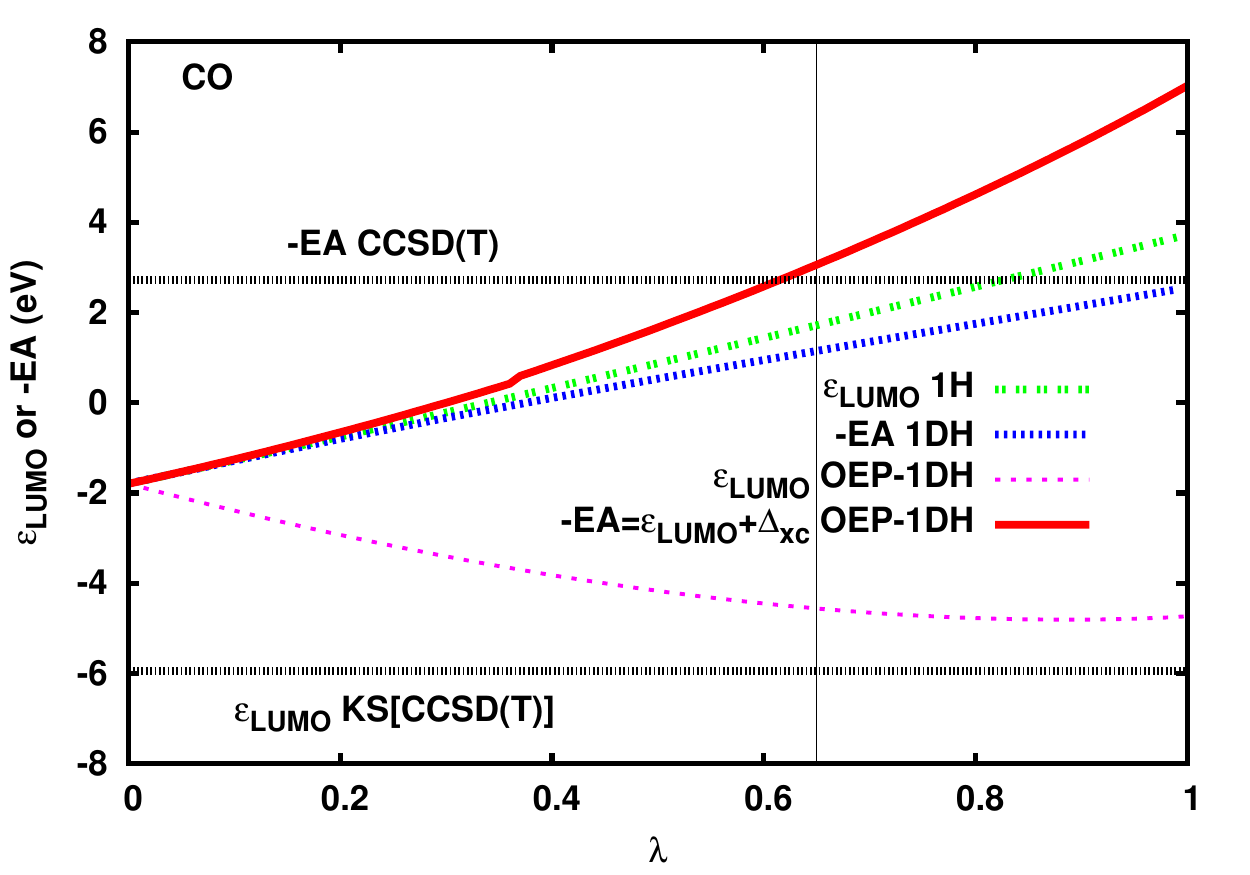}
\includegraphics[scale=0.45]{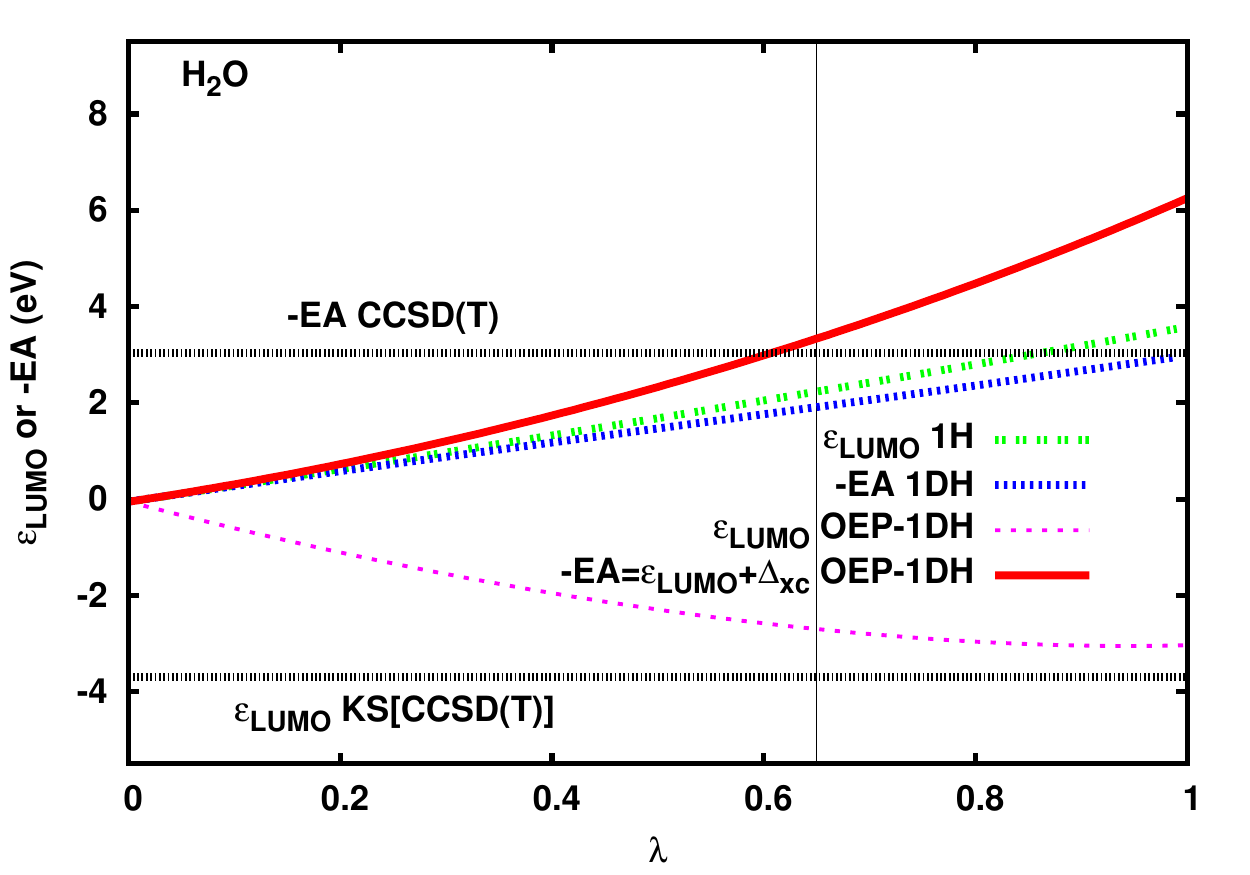}
\caption{LUMO orbital energies in the 1H [Eq.~(\ref{hnlpsi})] and OEP-1DH [Eq.~(\ref{hOEPpsi})] approximations and minus EAs in the 1DH [Eq.~(\ref{EA1DH})] and OEP-1DH [Eq.~(\ref{EAOEP1DH})] approximations using the BLYP functional as a function of $\lambda$. The reference EA values were calculated as CCSD(T) total energy differences with the same basis sets, and the reference HOMO energy values were calculated by KS inversion of the CCSD(T) densities. The vertical lines correspond to $\lambda=0.65$, i.e. the value recommended for 1DH with the BLYP functional in Ref.~\onlinecite{ShaTouSav-JCP-11}. For Be, the OEP-1DH calculations are unstable for $\lambda >0.86$. 
}
\label{fig:ea}
\end{figure*}

Figure~\ref{fig:ea} reports, for each system, the LUMO orbital energy in the 1H [Eq.~(\ref{hnlpsi})] and OEP-1DH [Eq.~(\ref{hOEPpsi})] approximations and minus the EAs in the 1DH [Eq.~(\ref{EA1DH})] and OEP-1DH [Eq.~(\ref{EAOEP1DH}), i.e. including the derivative discontinuity] approximations as a function of $\lambda$. The reference EAs are from CCSD(T) calculations with the same basis sets, whereas the reference KS LUMO orbital energies have been obtained by inversion of the KS equations using CCSD(T) densities as input. The reference $-$EAs are all positive for the systems considered, meaning that the anions are unstable. These positive values are an artifact of the incompleteness of the basis set. In a complete basis set, the EAs should be either negative (i.e., the anion is more stable than the neutral system) or zero (i.e., the anion dissociates into the neutral system and a free electron). Even though the reported EA values are thus not converged with respect to the basis set, the EAs given by different methods can nevertheless be compared for a fixed basis set. By contrast, the reference KS LUMO orbital energies are all correctly negative with the basis set employed. This is due to the fact that the KS LUMO does not represent a state with an additional electron but a bound excited state of the neutral system, which requires much less diffuse basis functions to describe. We note however that, in the case of the CO and H$_2$O molecules, the reported KS LUMO orbital energies are not well converged with respect to the basis set due to the lack of diffuse basis functions. Again, we can nevertheless meaningfully compare them with the reference data obtained with the same basis sets.

The LUMO orbital energy in the 1H approximation represents the simplest approximation to $-$EA (and not to the KS LUMO orbital energy since it is obtained with the nonlocal HF potential) available when doing a 1DH calculation. At $\lambda=0$, the 1H approximation reduces to standard KS with the BLYP functional, and we recover the fact that LUMO orbital energy with a semilocal DFA like BLYP is roughly half way between the exact KS LUMO orbital energy and the exact $-$EA (i.e., $\varepsilon_L^{\text{DFA}} \approx \varepsilon_L^{\text{exact}} +\Delta_\text{xc}/2$, see e.g. Ref.~\onlinecite{TeaDepToz-JCP-08}). At $\lambda=1$, the 1H approximation reduces to standard HF, and in this case the LUMO orbital energy is a quite good estimate of $-$EA (within the finite basis set) for the systems considered here. At $\lambda=0.65$, the 1H LUMO orbital energy underestimates $-$EA by about 0.25 to 1 eV, depending on the system.

The 1DH approximation gives $-$EAs rather close to the 1H ones, which indicates that the MP2 correlation term has only a modest effect on this quantity for the systems considered. Coming now to the OEP-1DH results, the LUMO orbital energy obtained in these calculations has a behavior as a function of $\lambda$ which is clearly distinct from the other curves. Starting from the BLYP LUMO orbital energy value at $\lambda=0$, it becomes more negative as $\lambda$ is increased, and becomes an increasingly accurate approximation to the exact KS LUMO orbital energy (and not to $-$EA). This is an essential difference between having a local EXX (and GL2) potential instead of a nonlocal HF potential. At $\lambda=0.65$, the OEP-1DH LUMO orbital energies underestimate the reference KS LUMO energies by about 1 to 2 eV. The estimate of $-$EA in the OEP-1DH approximation is obtained by adding the derivative discontinuity $\Delta_\text{xc}$ to the OEP-1DH LUMO orbital energy. For the closed-shell systems considered here, the derivative discontinuity is largely dominated by the exchange contribution. The derivative discontinuity is systematically overestimated in OEP-GL2, leading to $-$EAs that are much too high. It turns out that, at the recommended value $\lambda=0.65$, OEP-1DH gives $-$EAs which agree with the reference values within 0.4 eV for the systems considered. Thus, the OEP-based self-consistency improves the accuracy of EAs.

\subsection{Exchange-correlation and correlation potentials}

\begin{figure*}
\includegraphics[scale=0.45]{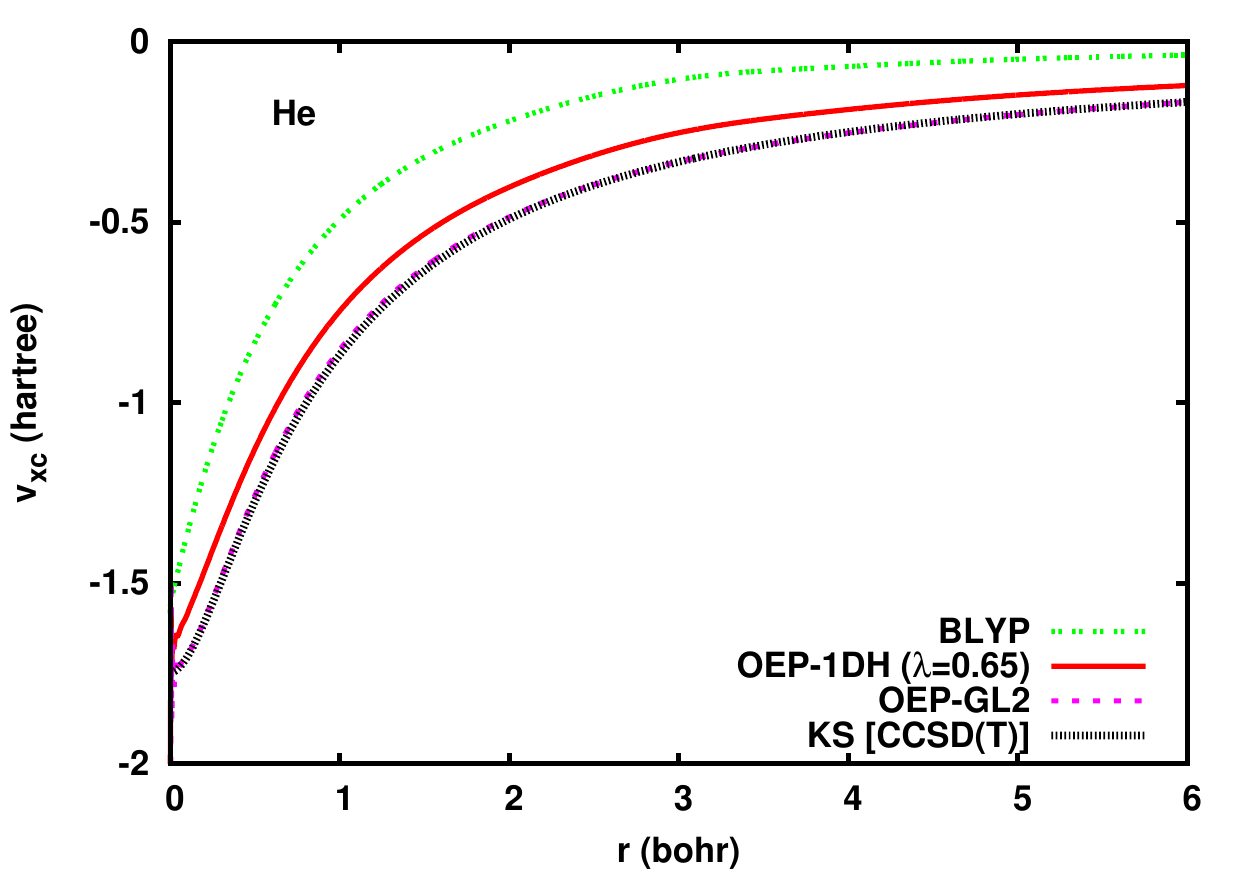}
\includegraphics[scale=0.45]{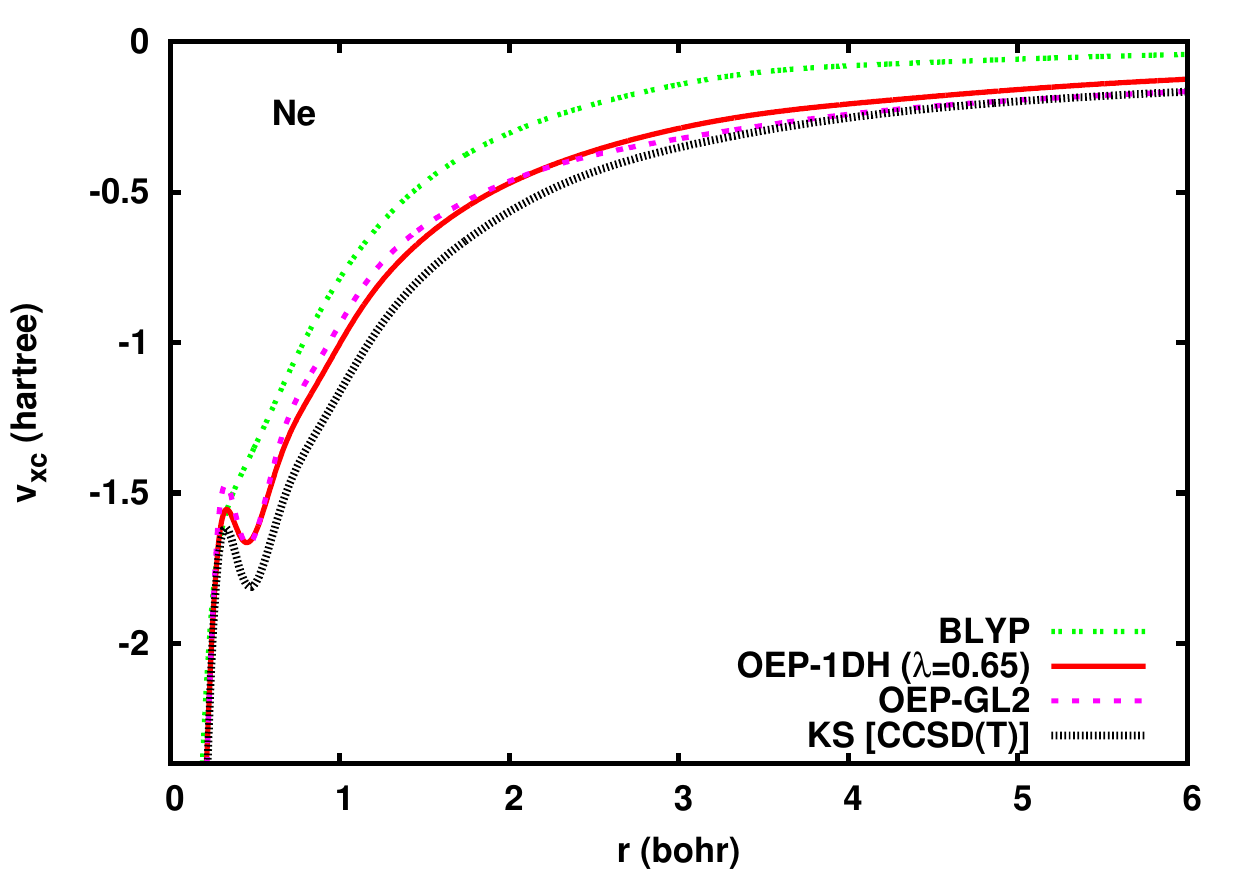}
\includegraphics[scale=0.45]{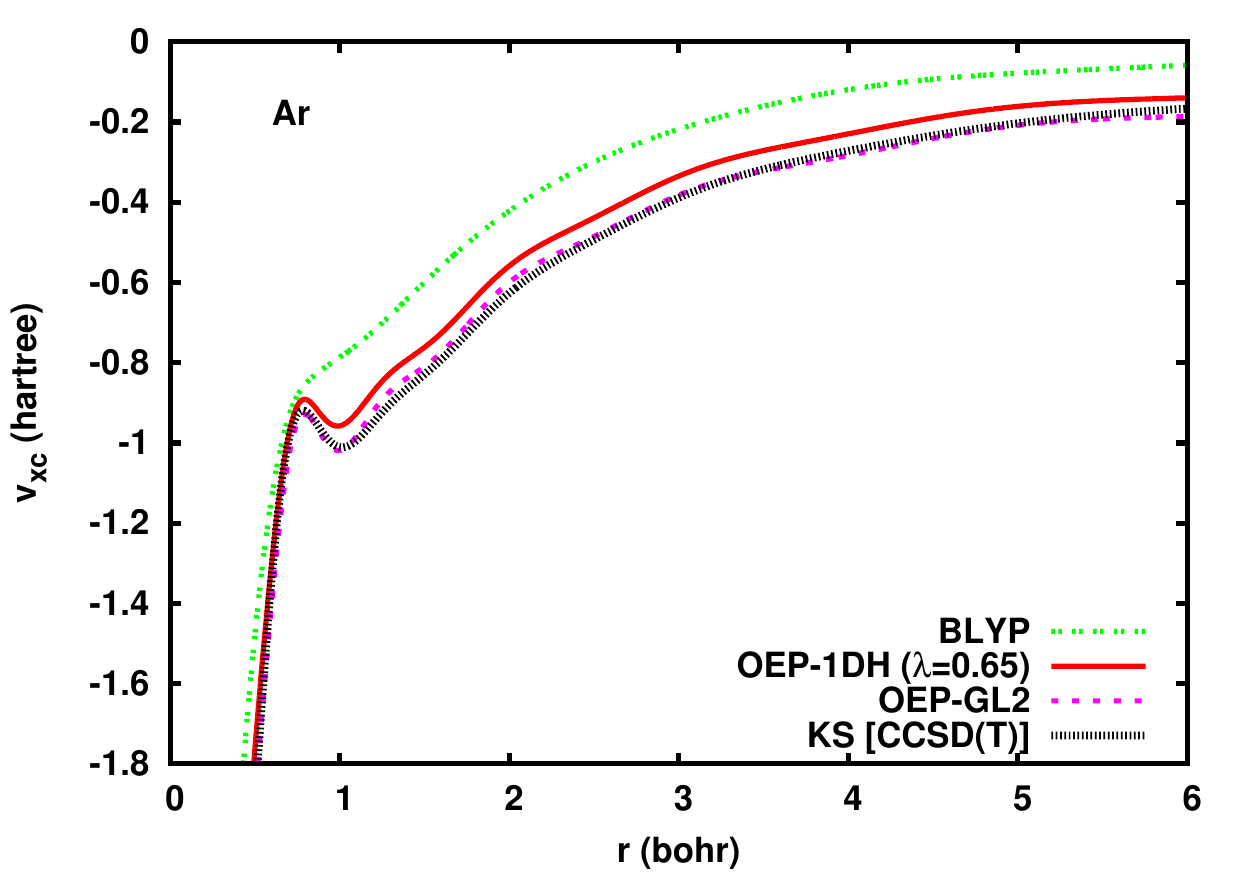}
\includegraphics[scale=0.45]{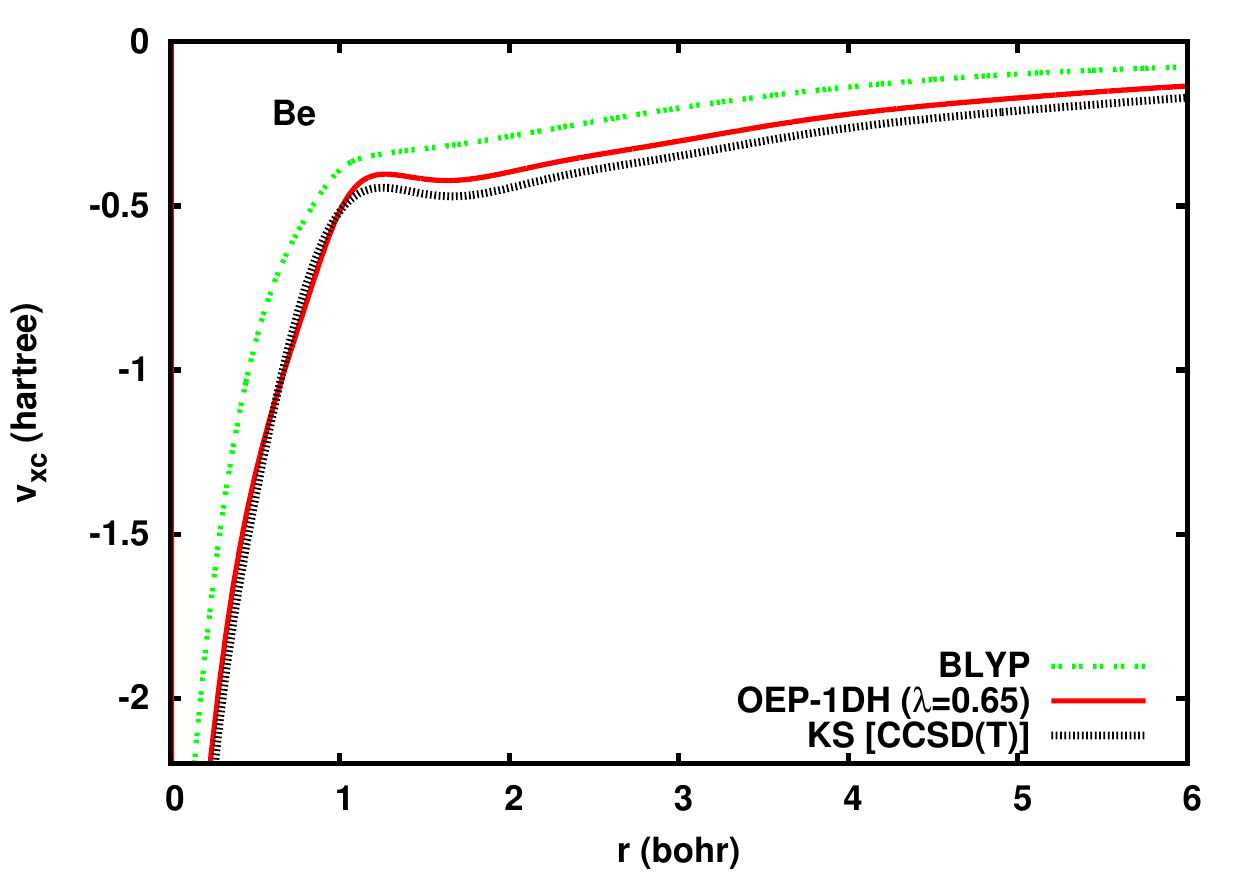}
\includegraphics[scale=0.45]{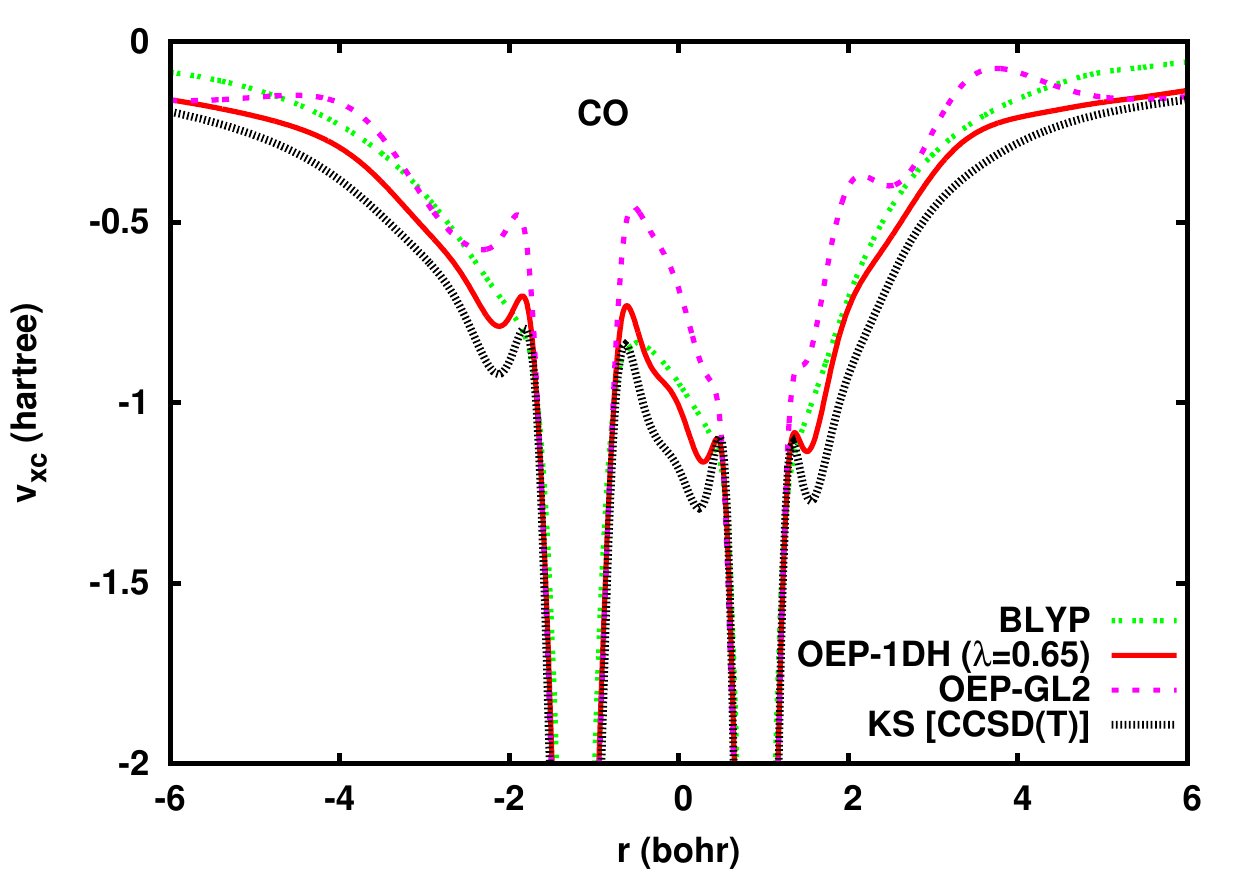}
\includegraphics[scale=0.45]{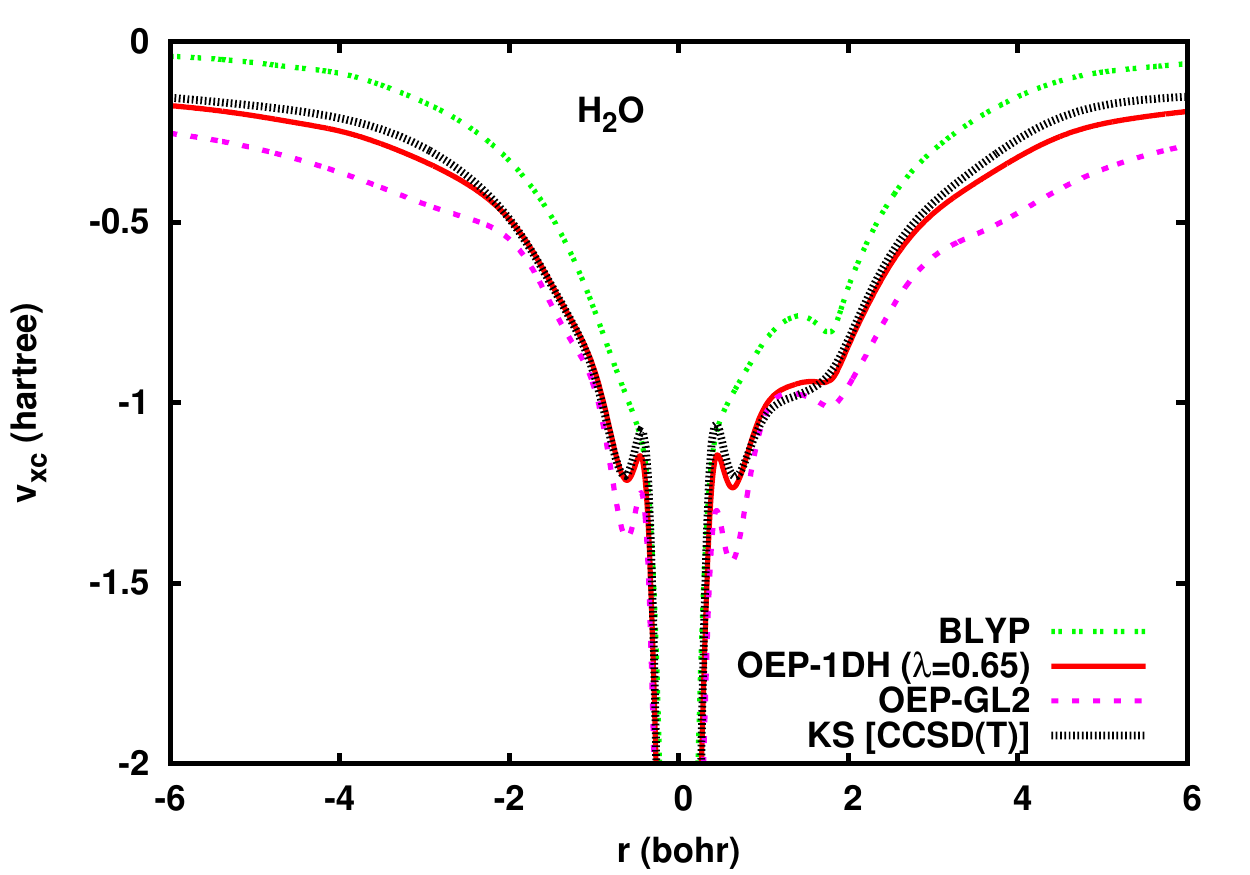}
\caption{Exchange-correlation potentials calculated with the OEP-1DH approximation [Eq.~(\ref{vxcOEP1DH})] using the BLYP functional at the recommended value $\lambda=0.65$, and at the extreme values $\l=0$ (standard BLYP) and $\l=1$ (OEP-GL2). The reference potentials were calculated by KS inversion of the CCSD(T) densities. For Be, the OEP-GL2 calculations are unstable. For CO, the potential is plotted along the direction of the bond with the C nucleus at $-1.21790$ bohr and the O nucleus at $0.91371$ bohr. For H$_2$O, the potential is plotted along the direction of a OH bond with the O nucleus at $0.0$ and the H nucleus at $1.81225$ bohr.} 
\label{fig:vxc}
\end{figure*}

Figure~\ref{fig:vxc} shows the exchange-correlation potentials calculated by OEP-1DH at the recommended value of $\lambda=0.65$, as well as the potentials obtained at the extreme values of $\lambda$, corresponding to KS BLYP ($\lambda=0$) and OEP-GL2 ($\lambda=1$). The reference potentials have been obtained by employing the KS inversion approach using the CCSD(T) densities. 

The BLYP exchange-correlation potentials are not negative enough, they do not describe well the shell structure (core/valence transition), and decay too fast at large distances. The OEP-GL2 exchange-correlation potentials have the correct $-1/r$ asymptotic behavior and are quite accurate for the rare-gas atoms (especially for He and Ar), but have too much structure for CO and H$_2$O. For Be, the OEP-GL2 calculation is unstable. The OEP-1DH exchange-correlation potentials do not have quite the correct asymptotic behavior since they decay as $-\lambda/r$, but for $\lambda=0.65$ they have reasonable shapes in the physically relevant region of space. Note in particular that the OEP-1DH calculation yields a stable solution for Be. For CO and H$_2$O, the OEP-1DH exchange-correlation potentials actually improve over both the BLYP and OEP-GL2 exchange-correlation potentials. Therefore, even though the recommended value of $\lambda=0.65$ was determined based on energetical properties of the standard non-self-consistent 1DH scheme, it appears that this value of $\lambda$ also gives reasonable exchange-correlation potentials as well.

\begin{figure*}
\includegraphics[scale=0.45]{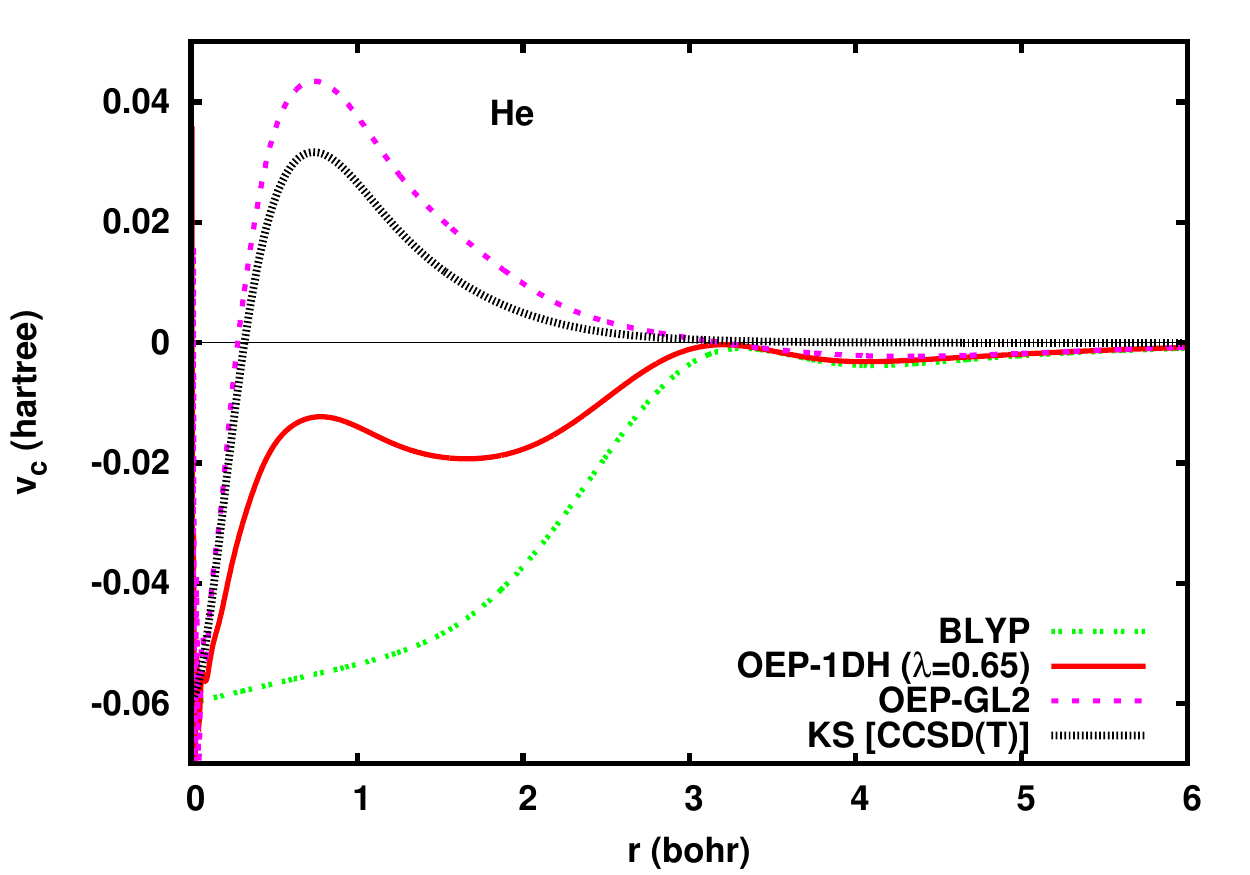}
\includegraphics[scale=0.45]{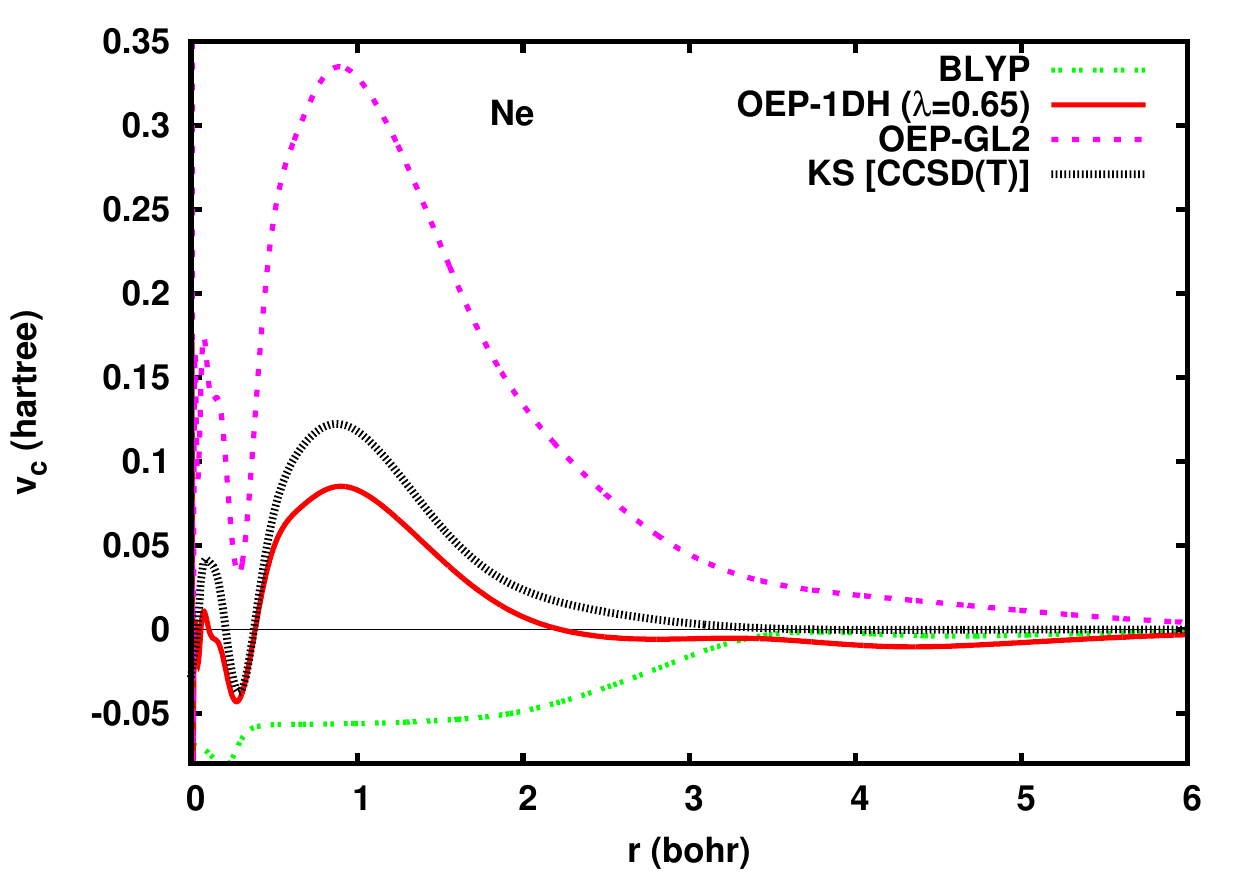}
\includegraphics[scale=0.45]{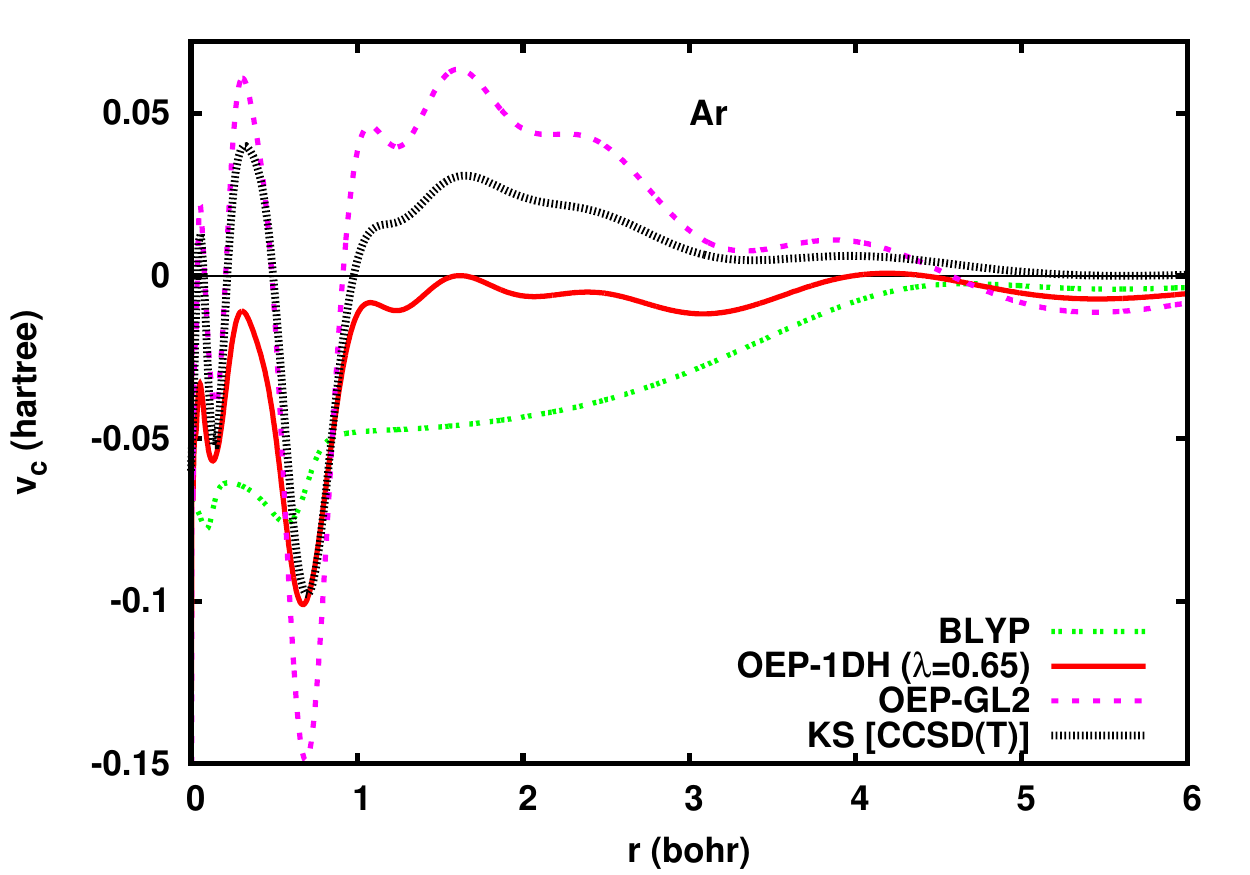}
\includegraphics[scale=0.45]{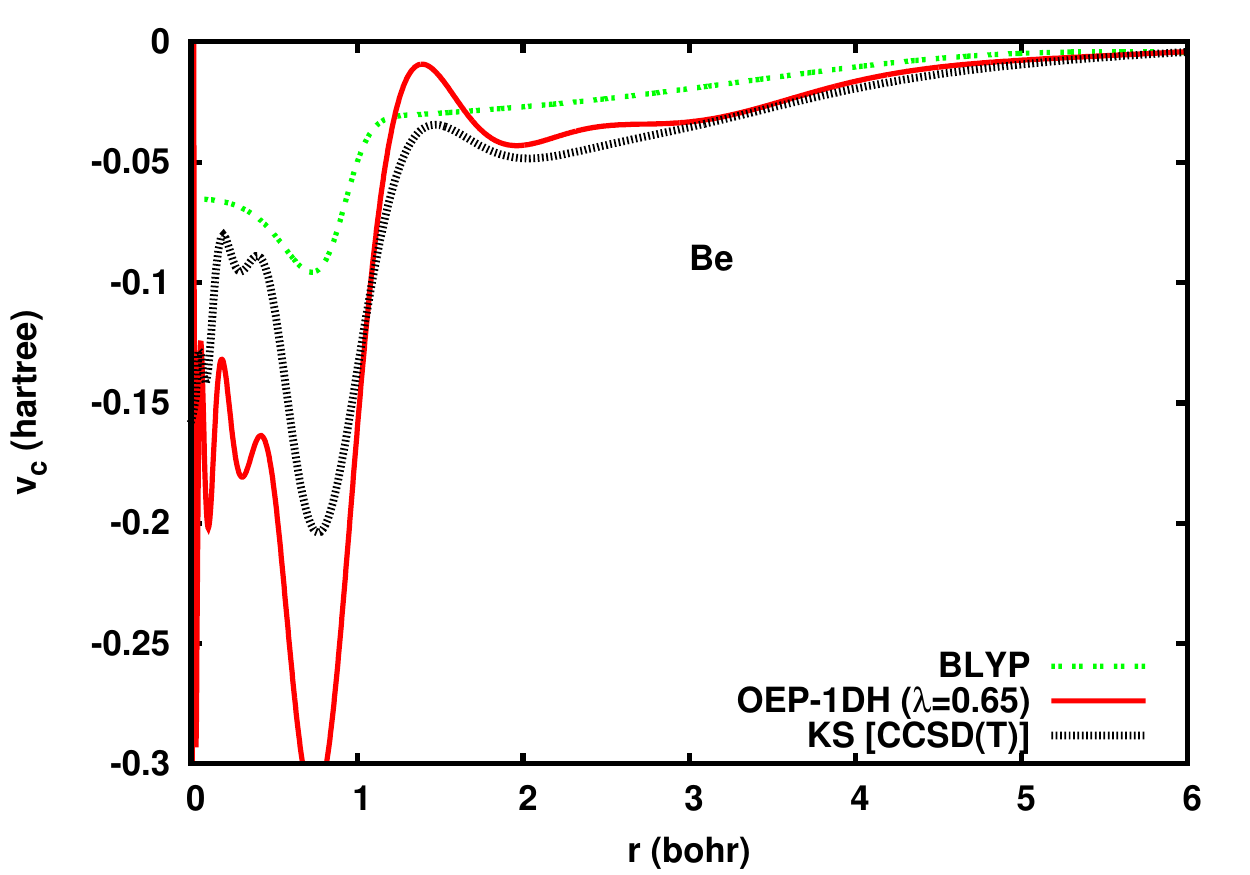}
\includegraphics[scale=0.45]{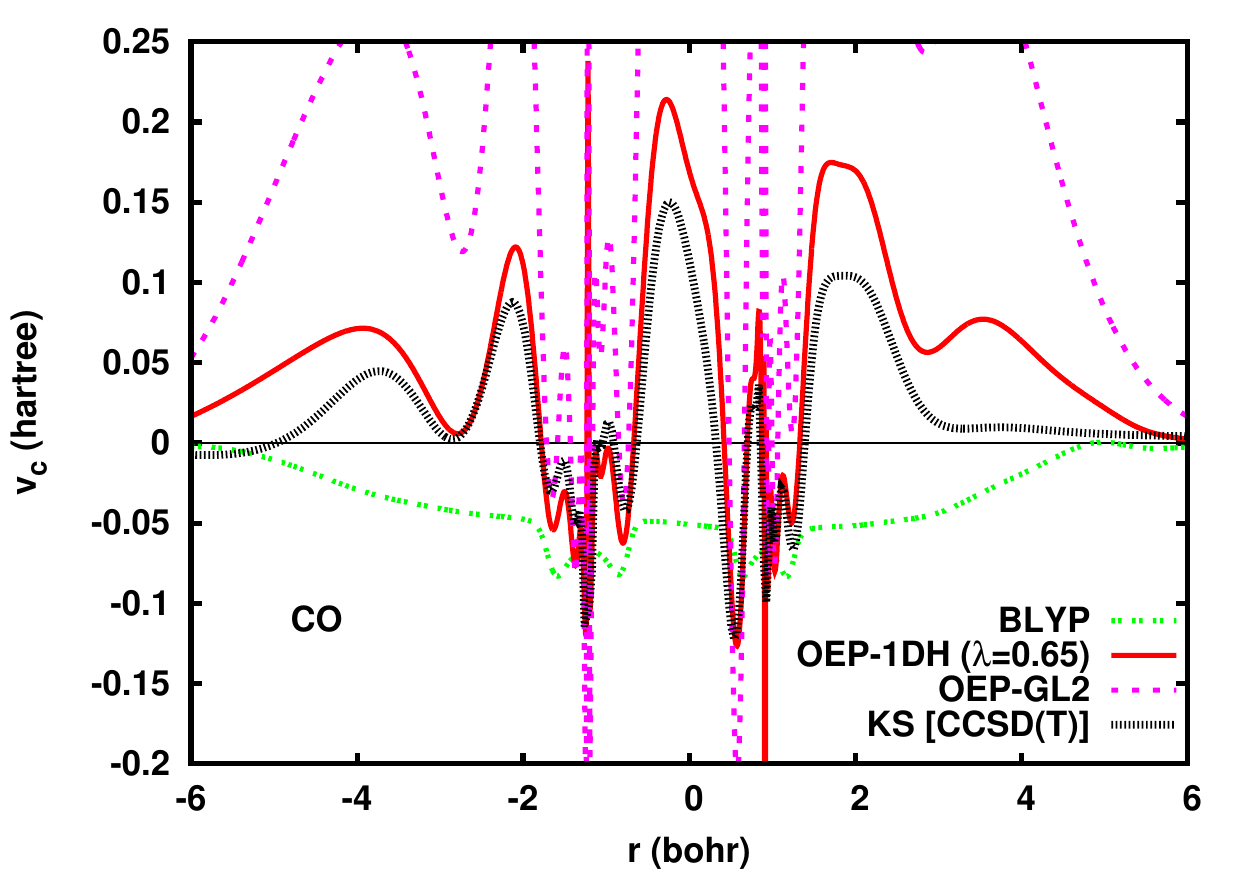}
\includegraphics[scale=0.45]{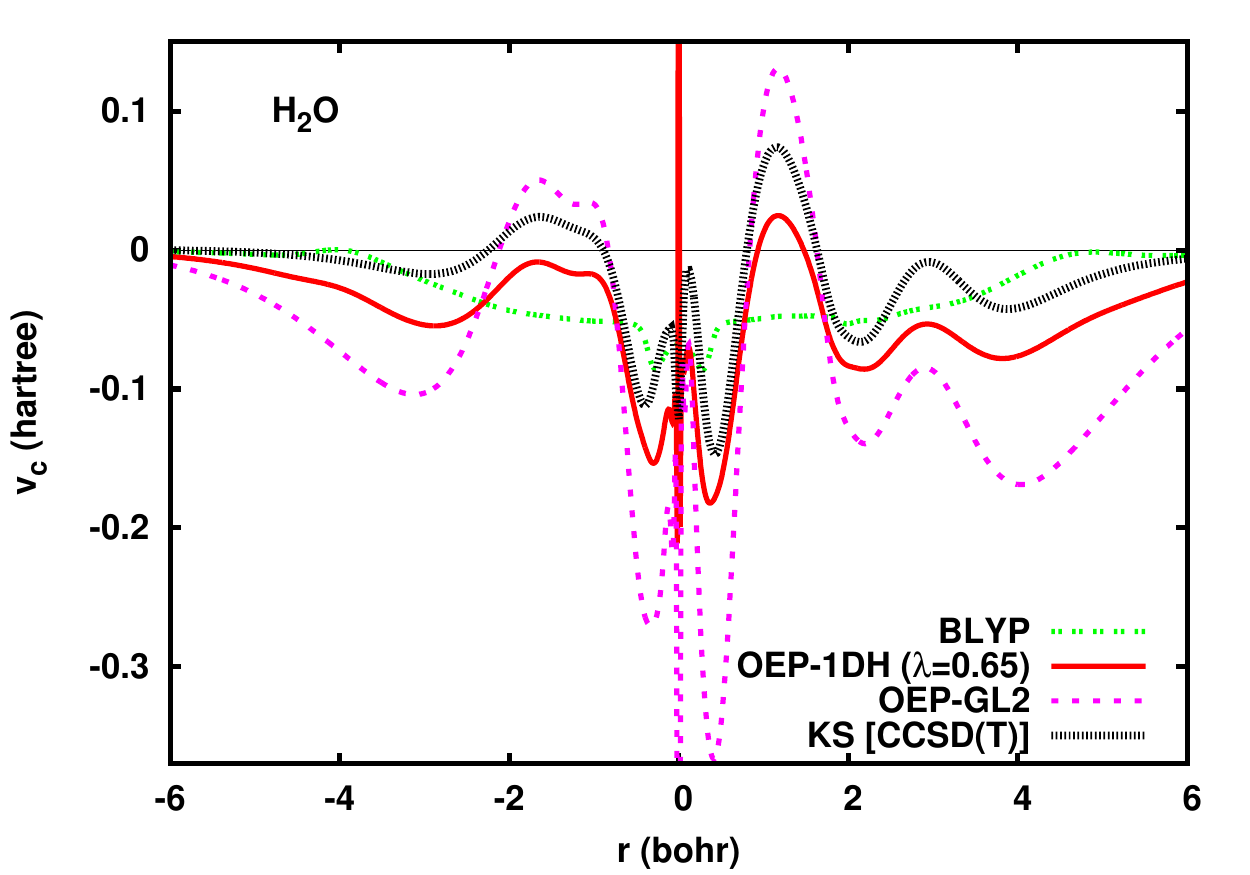}
\caption{Correlation potentials calculated with the OEP-1DH approximation [correlation terms in Eq.~(\ref{vxcOEP1DH})] using the BLYP functional at the recommended value $\lambda=0.65$, and at the extreme values $\l=0$ (standard BLYP) and $\l=1$ (OEP-GL2). The reference potentials were calculated by KS inversion of the CCSD(T) densities. For Be, the OEP-GL2 calculation is unstable. For CO, the potential is plotted along the direction of the bond with the C nucleus at $-1.21790$ bohr and the O nucleus at $0.91371$ bohr. For H$_2$O, the potential is plotted along the direction of a OH bond with the O nucleus at $0.0$ and the H nucleus at $1.81225$ bohr.}
\label{fig:vc}
\end{figure*}

The correlation part of the potentials are plotted in Figure~\ref{fig:vc}. The correlation potentials in BLYP calculations (i.e., the LYP correlation potential evaluated at the self-consistent BLYP density) are unable to reproduce the complex structure of the reference correlation potentials. Note that this is in spite of the fact that LYP correlation energies are usually reasonably accurate. On the contrary, the OEP-GL2 correlation potentials tend to be largely overestimated, as previously observed~\cite{GraTeaSmiBar-JCP-11,GraFabTeaSmiBukDel-JCP-14}. Overall, the OEP-1DH correlation potentials at $\lambda=0.65$ have fairly reasonable shapes, providing a good compromise between the understructured BLYP and the overestimated OEP-GL2 correlation potentials.

\subsection{Correlated densities}

\begin{figure*}
\includegraphics[scale=0.45]{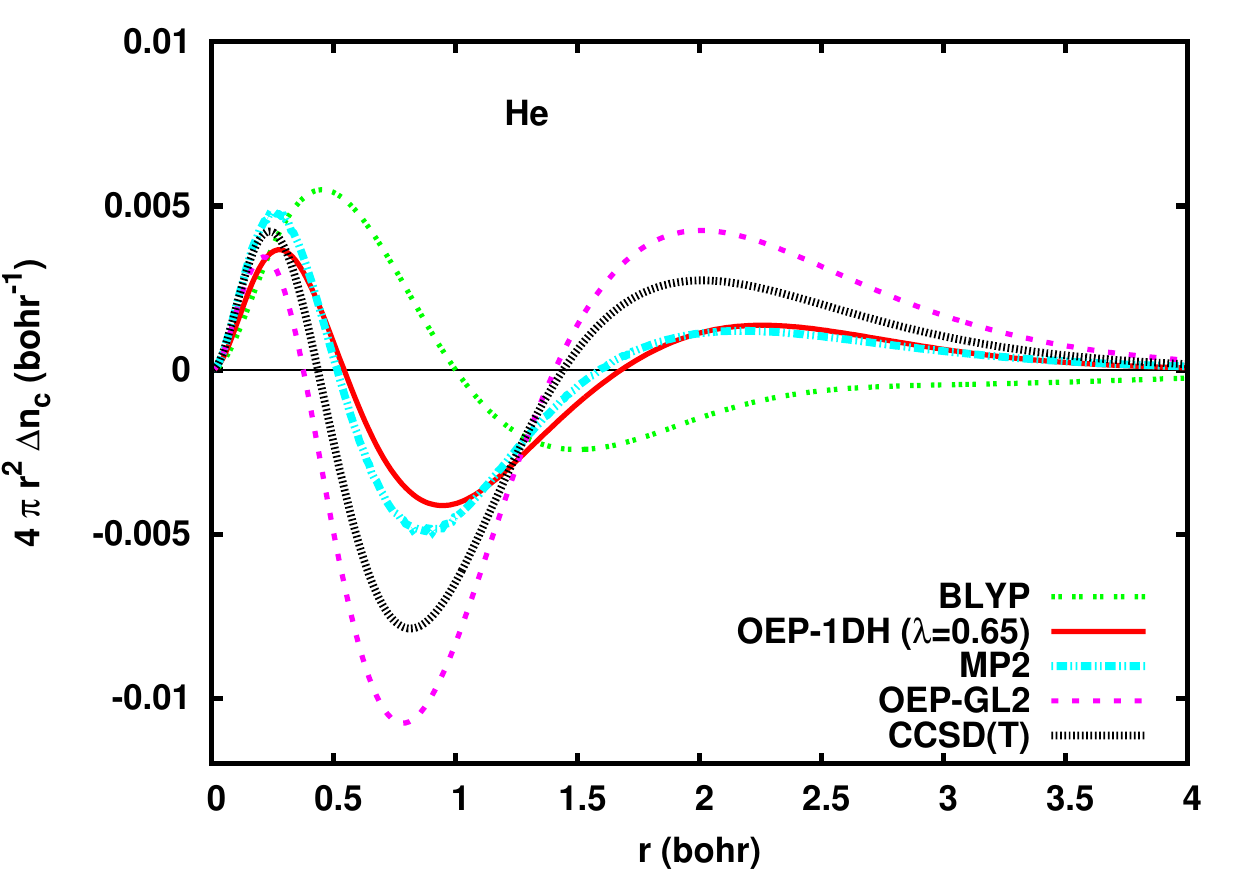}
\includegraphics[scale=0.45]{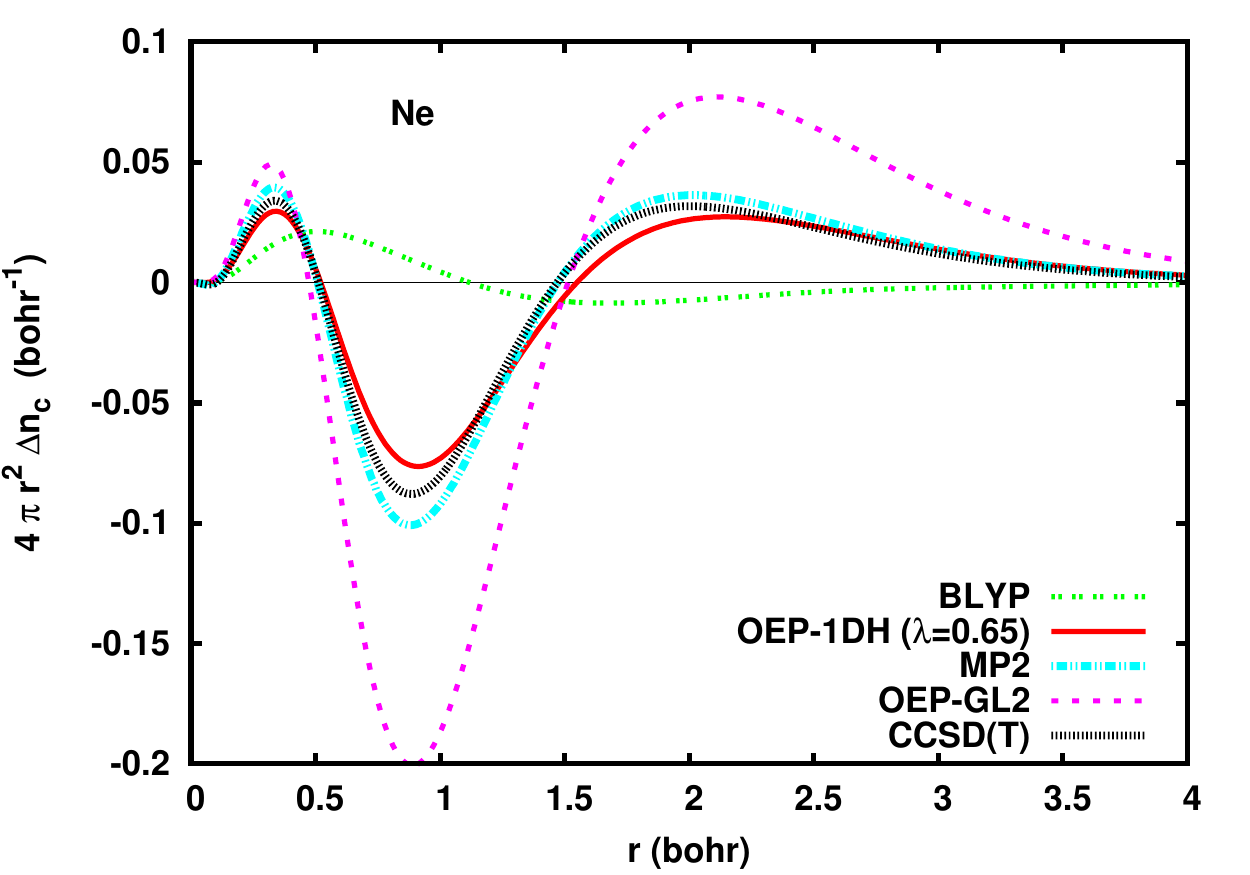}
\includegraphics[scale=0.45]{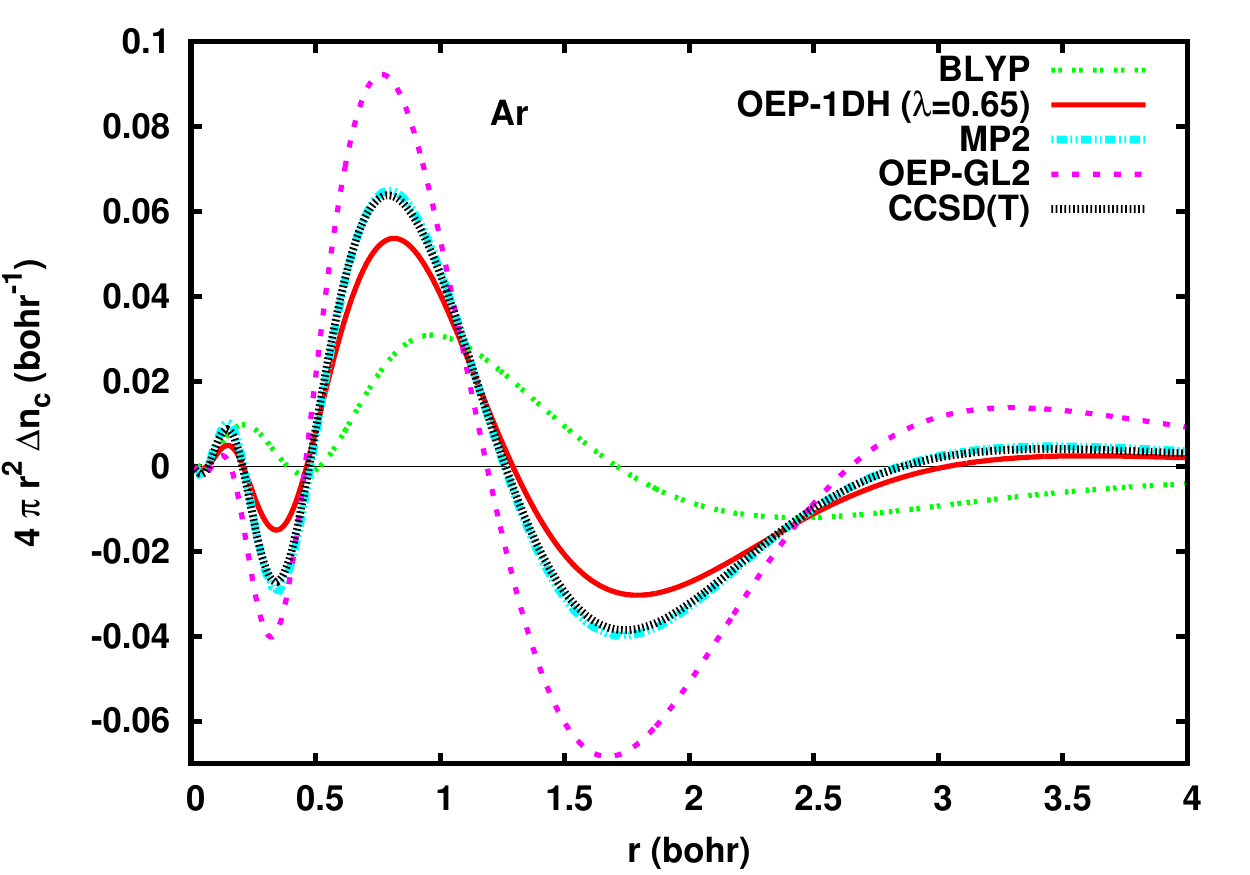}
\includegraphics[scale=0.45]{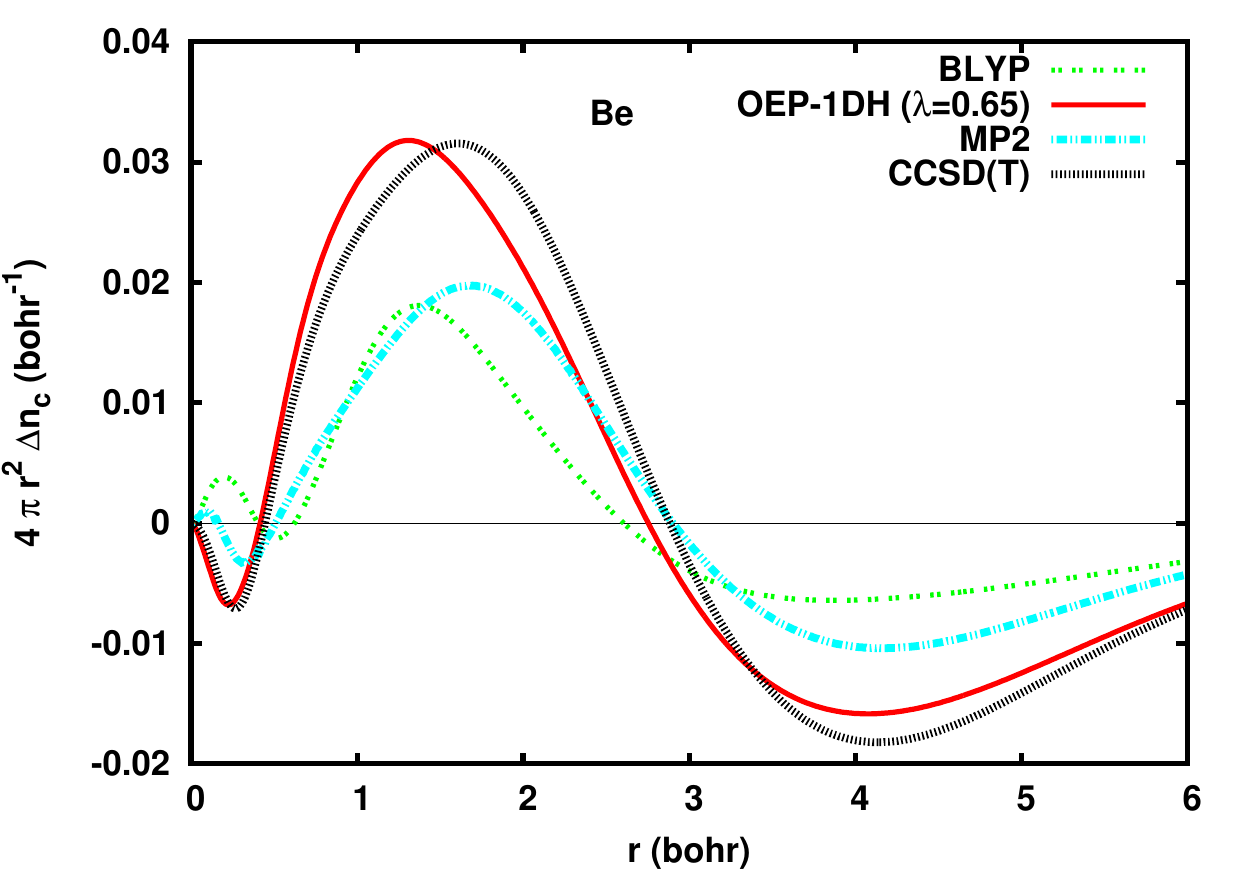}
\includegraphics[scale=0.45]{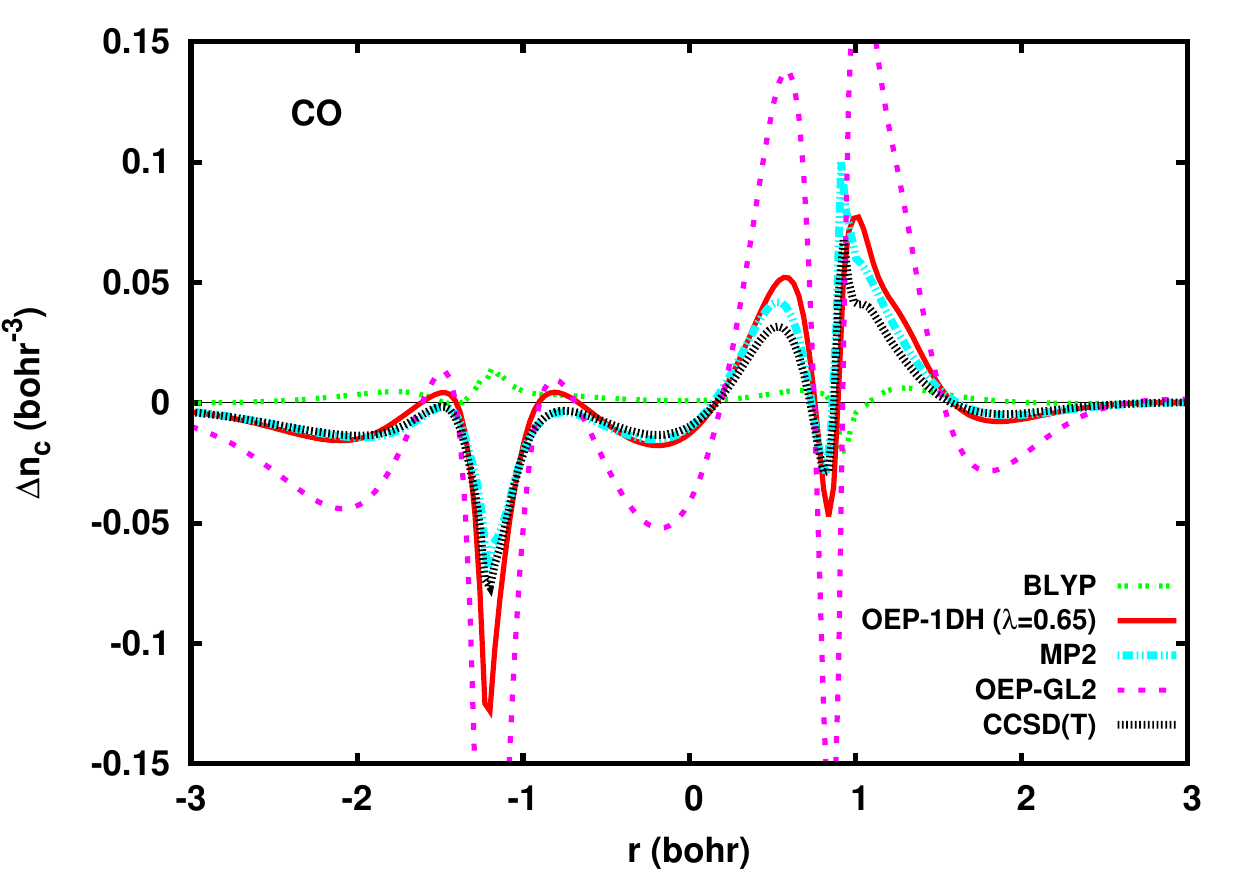}
\includegraphics[scale=0.45]{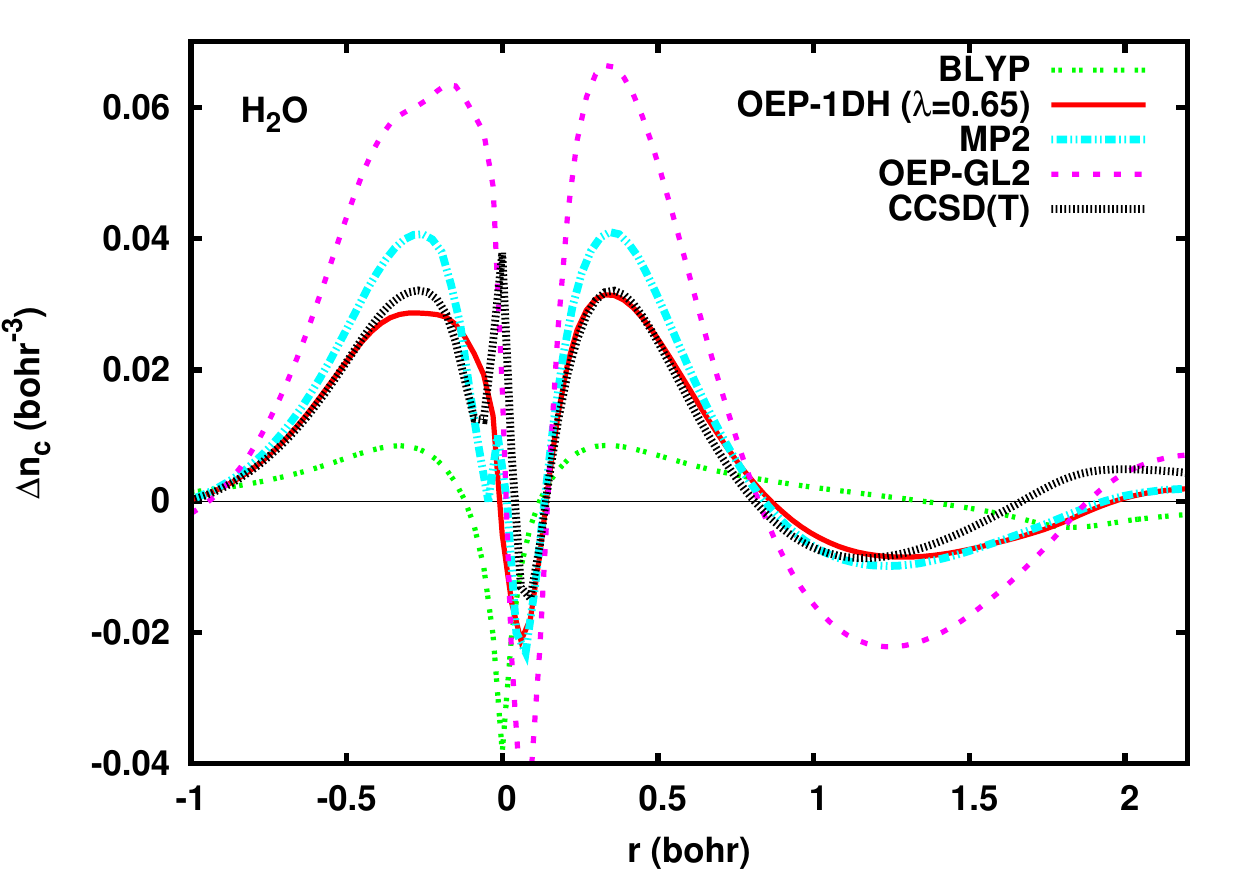}
\caption{Correlated density calculated with the OEP-1DH approximation using the BLYP functional at the recommended value $\lambda=0.65$, and at the extreme values $\l=0$ (standard BLYP) and $\l=1$ (OEP-GL2). The reference correlated densities are calculated with CCSD(T). For comparison, correlated densities calculated with standard MP2 are also shown. For Be, the OEP-GL2 calculation is unstable. For CO, the potential is plotted along the direction of the bond with the C nucleus at $-1.21790$ bohr and the O nucleus at $0.91371$ bohr. For H$_2$O, the potential is plotted along the direction of a OH bond with the O nucleus at $0.0$ and the H nucleus at $1.81225$ bohr.}
\label{fig:density}
\end{figure*}

The analysis of the correlated densities provides a useful tool for the detailed examination of the correlation effects on the electronic density and for the test of exchange-correlation approximations in DFT~\cite{Jankowski:2009:DRD,Jankowski:2010:DRD,GraTeaSmiBar-JCP-11,GraTeaFabSmiBukDel-MP-14,Smiga2014125,Buksztel2016}. Thus, in Figure~\ref{fig:density} we report correlated densities calculated by OEP-1DH at the recommended value of $\lambda=0.65$, as well as the correlated densities obtained at the extreme values of $\lambda$, corresponding to KS BLYP ($\lambda=0$) and OEP-GL2 ($\lambda=1$). The correlated density is defined as $\Delta n_\text{c} (\b{r}) = n(\b{r}) - n_\text{x-only}(\b{r})$ where $n(\b{r})$ is the total density calculated with the full exchange and correlation terms and $n_\text{x-only}(\b{r})$ is the density calculated with only the exchange terms (see Refs.~\onlinecite{Jankowski:2009:DRD,Jankowski:2010:DRD,GraTeaSmiBar-JCP-11,GraTeaFabSmiBukDel-MP-14} for discussions on different definitions of correlated densities). The reference correlated densities have been calculated with the CCSD(T) method as $\Delta n_\text{c,CCSD(T)} (\b{r}) = n_\text{CCSD(T)}(\b{r}) - n_\text{HF}(\b{r})$, while $n_\text{CCSD(T)}(\b{r})$ was obtained from the CCSD(T) relaxed density matrix \cite{hansch:1984:zvec,ricamo:1985:relden,bartlett:1989:relden} constructed using the Lagrangian approach \cite{jorg:1988:lag,koch:1990:lag,hald:2003:lag}. For comparison, correlated densities calculated in standard MP2 (1DH with $\lambda=1$) with the same relaxed density-matrix approach are also shown.
Due to our current implementation limitations (lack of the relaxed density-matrix approach for the 1DH approximation), the correlated densities for the non-self-consistent 1DH approximation have not been calculated.

At $\lambda=0$, i.e. in KS BLYP calculations, the correlated densities are mostly very much underestimated. At $\lambda=1$, the correlated densities are largely overestimated with OEP-GL2. At $\lambda=0.65$, the OEP-1DH correlated densities tend to be quite accurate, achieving a good balance between the underestimated BLYP correlated densities at $\lambda=0$ and the overestimated OEP-GL2 correlated densities at $\lambda=1$. The OEP-1DH correlated densities are overall similar in accuracy to the MP2 and CCSD(T) correlated densities.

\section{Conclusion}
\label{sec:conclusion}

In this work, we have proposed an OEP-based self-consistent DH scheme in which the orbitals are optimized with a local potential including the MP2 correlation contribution. While staying in the philosophy of the KS scheme with a local potential, this scheme constitutes an alternative to the orbital-optimized DH scheme of Peverati and Head-Gordon~\cite{PevHea-JCP-13}. 

We have implemented a one-parameter version of this OEP-based self-consistent DH scheme using the BLYP density-functional approximation and compared it to the corresponding non-self-consistent DH scheme for calculations on a few closed-shell atoms and molecules. While the OEP-based self-consistency does not provide any improvement for the calculations of ground-state total energies and ionization potentials, it does improve the accuracy of electron affinities and restores the meaning of the LUMO orbital energy as being connected to a neutral excitation energy. Moreover, the OEP-based self-consistent DH scheme provides reasonably accurate exchange-correlation potentials and correlated densities. In comparison to the standard OEP-GL2 method~\cite{GorLev-PRA-94,GorLev-IJQC-95,GraHirIvaBar-JCP-02,BarGraHirIva-JCP-05,MorWuYan-JCP-05}, our OEP-based self-consistent DH scheme is more stable and removes the large overestimation of correlation effects.

Additional work can be foreseen to exploit the full power of the OEP-based self-consistent DH scheme. For example, our scheme should be tested against more systems, including open-shell ones for which we expect the OEP self-consistency to provide advantages similar to the orbital-optimized DH scheme of Ref.~\onlinecite{PevHea-JCP-13}. It would also be interesting to apply linear-response time-dependent DFT on our OEP-based self-consistent DH scheme to calculate excitation energies. Finally, the present procedure should be applied to the range-separated DH approach~\cite{AngGerSavTou-PRA-05} which has the advantage of having a fast basis convergence~\cite{FraMusLupTou-JCP-15}.

\section*{Acknowledgements}
We thank Andreas Savin for discussions, as well as Andrew Teale for providing the reference 
data for the exchange and exchange-correlation potentials.  This work was supported by French 
state funds managed by CALSIMLAB, within the Investissements d'Avenir program under reference ANR-11-IDEX-0004-02, by the Polish-French
bilateral programme POLONIUM, and by the Polish National Science Center under Grant No. DEC-2013/11/B/ST4/00771.


%
   \end{document}